\documentclass[aip,rsi,amsmath,amssymb,reprint]{revtex4-2}
\pdfoutput=1
\usepackage{graphicx}
\usepackage{dcolumn}
\usepackage{bm}
\usepackage[utf8]{inputenc}
\usepackage[T1]{fontenc}

\usepackage{mathptmx}
\usepackage{siunitx}
\usepackage{graphicx}
\usepackage{dcolumn}
\usepackage{braket}
\usepackage{xfrac}
\usepackage{hyperref}
\usepackage{tikz-cd}
\usepackage{float}
\usepackage{amsmath}
\usepackage[nameinlink,capitalise]{cleveref}
\usepackage{mathtools}

\usepackage{bm}
\DeclareSIUnit\gauss{G}
\DeclareSIUnit{\au}{{a.u.}}
\hypersetup{
colorlinks=true,
linkcolor=blue,
citecolor=blue,
filecolor=green,
urlcolor=blue,
}

\begin{document}


\title{The role of low energy resonances in the stereodynamics of cold He+D$_2$ collisions}

\author{Pablo G. Jambrina}
\affiliation{Departamento de Qu\'imica F\'isica, University of Salamanca, Salamanca
37008, Spain.}
\email{pjambrina@usal.es}
\author{Masato Morita}
\affiliation{Department of Chemistry and Biochemistry, University of Nevada, Las Vegas, Nevada 89154, USA.}%
\author{James F. E. Croft}
\affiliation{Department of Physics, University of Otago, Dunedin
9054, New Zealand.}
\affiliation{Dodd-Walls Centre for Photonic and Quantum
Technologies, Dunedin 9054, New Zealand.}
\email{j.croft@otago.ac.nz}
\author{F. Javier Aoiz}
\affiliation{Departamento de Qu\'imica F\'isica, Universidad Complutense, Madrid
28040, Spain.}
\email{aoiz@quim.ucm.es}
\author{Naduvalath Balakrishnan}%
\email{naduvala@unlv.nevada.edu}
\affiliation{Department of Chemistry and Biochemistry, University of Nevada, Las Vegas, Nevada 89154, USA.}%

\date{\today}

\begin{abstract}

In recent experiments using the Stark-induced Adiabatic Raman Passage (SARP) technique, Zhou et al. measured the product's angular distribution for the collisions between He and aligned D$_2$ molecules at cold collision energies. The signatures of the angular distributions were attributed to a
$\ell$=2 resonance that  governs scattering at low energies. A first principles quantum
mechanical treatment of this problem is presented here using a highly accurate interaction
potential for the He-H$_2$ system. Instead, our results predict a very intense $\ell$=1 resonance at low
energies, leading to angular distributions that differ from those measured in the experiment. A good agreement with the experiment is achieved only when the $\ell$=1 resonance is
artificially removed, for example, by excluding the lowest energies present in the experimental
velocity distribution. Our analysis revealed that neither the position nor the intensity of the $\ell$=1 resonance significantly changes when the interaction potential is modified within its predicted uncertainties.  Energy-resolved measurements may help to resolve the discrepancy.

\end{abstract}

\maketitle

In molecular scattering, initial collision conditions that aid the system to reach the transition
state geometries lead to higher reaction yields, while conditions that impede the system to reach
such geometry lead to lower yields. This statement is usually related to the control of chemical
reactions but it can be generalized to inelastic collisions.

Experiments that study the outcome of a molecular collision depending on the initial conditions
have flourished in the last 10 years (see for example Refs.
\citenum{Wang2011,Wang2012,Wang2016a,Wang2016b,Brouard2015,Chadwick2014,Brouard2013,VKKBOAGM:NC18,OGVAKNAGBM:NC17,SLLMJACC:NC18,2017_Science_Perreault,2018_NatChem_Perreault,sarp_hed2,HWBJA:NC19,CPWJAB:JPCL21,HBWGJAB:PCCP20,WHJAB:JPCA19,sarp_hed2,sarp_hed2_science,SARP_HD-He}).
Besides their importance for elucidating collision mechanisms, these experiments constitute a
particularly effective probe of ab initio electronic-structure calculations and scattering methods.
For inelastic collisions of NO(A$^2\Sigma$) with He, Chandler, Costen, and coworkers determined how the
orientation of the product's angular momentum ($\bm{j'}$) depends on the initial orientation of the
NO molecule. \cite{SLLMJACC:NC18} In Brouard's group, differences in the differential cross
sections (DCS) for collisions between Rare Gases (Rg) and NO(X$^2\Pi$) molecules  were observed depending on
the NO orientation: whether the collision was head-on or side-on and whether the Rg hits the
molecule close to the O or N atom. \cite{CPWJAB:JPCL21,HBWGJAB:PCCP20,WHJAB:JPCA19} Zare and
coworkers exploited the combination of co-propagating molecular beams with the Stark-induced adiabatic
Raman Passage (SARP) method for state-preparation and alignment of the molecules.
~\cite{2017_Science_Perreault,2018_NatChem_Perreault,SARP_HD-He,sarp_hed2,sarp_hed2_science,Frontiers21Zare}
In these experiments, the colliding molecules are co-propagated in the same molecular beam allowing
relative collision energies near 1 K though the  molecular velocities in the laboratory frame are
much higher (about 2000 m/s). This is a particularly interesting regime which when combined with the
SARP method allows for the study of sterodynamical preferences where only a few partial waves
contribute. As such this combination is a powerful tool to probe the interaction potential.~\cite{2017_Science_Perreault}

On the computational side, scattering calculations have been carried out to study how the outcome
of bimolecular collisions depends on the initial collision conditions at low energies. In
particular, since it has been experimentally determined that at cold energies  the excitation
function (cross section as a function of the collision energy) is governed by the presence of sharp
resonance peaks,\cite{Meeraker:S15,Meeraker:S20} a lot of effort has been devoted to establish to
what extent resonance peaks can be controlled by selective experimental preparations.
~\cite{2018_PRL_Croft,2019_JCP_Croft,2019_PRL_Jambrina,Morita_He-HD,morita_hcl-h2,Jambrina_PCCP_2021}

Recently, Zhou et al. presented a scattering experiment analogous to the double-slit experiment
involving He+D$_2(v=2,j=2)\to$ He+D$_2(v'=2,j'=0)$ inelastic collisions near 1 K that yielded an
interference term arising from two simultaneous bond-axis alignments of the D$_2$ molecule at $\pm$
45$^\circ$ relative to the SARP laser polarization.~\cite{sarp_hed2,sarp_hed2_science} The angular
distributions (DCS) for the inelastically scattered D$_2$ were found to be markedly different from
a uni-axial preparation at $45^\circ$ (or $135^\circ$) that does not include an interference term.
In this Letter, we provide a first principles analysis of the experimental results of Zhou et al.
using a highly accurate interaction potential for the He-H$_2$ system~\cite{BSP3} that was
benchmarked against high-resolution cavity measurements of line-shape parameters of H$_2$ perturbed
by helium~\cite{BSP3_2020_PRA} and rotational Raman spectrum of D$_2$ in He.~\cite{Martinez2018}
Our quantum  calculations reveal that scattering is governed by an $\ell$=1 resonance at collision
energies between 1 mK and 1\,K.  Striking differences between computational and experimental DCSs
are observed, and calculations can only reproduce the experimental results if the $\ell$=1
resonance is removed, which can be artificially done by excluding the low energies from our
calculations. We find that the theoretical results are largely insensitive to the choice of quantum
scattering method used and to  modifications of the He--D$_2$ interaction potential within the
uncertainties of the ab initio electronic structure calculations. Our results indicate the need for
energy resolved measurements to resolve the discrepancy between theory and experiment for this
benchmark system which may be accomplished using merged beam techniques.

%
Time independent quantum mechanical calculations were carried out using the coupled-channel (CC)
formalism~~\cite{1952_Blatt_Biedenharn_Rev,1960_Dalgarno_Arthurs,1977_Alexander_1,1977_Alexander_2}
as implemented in the MOLSCAT code.~\cite{MOLSCAT} Similar results were obtained using the ABC
quantum scattering code~\cite{ABC_code} that uses the CC method in hyperspherical coordinates. The
CC equations are constructed with a basis set that includes vibrational levels $v=0-3$ and
rotational levels $j=0-2$ for $v=3$, $j=0-8$ for $v=2$, $j=0-12$ for $v=1$, and $j=0-18$ for $v=0$.
The CC equations are propagated from an atom-molecule separation of $R_{\min}=2.0$ $a_0$ to
$R_{\max}=50-100$ $a_0$ (depending on the collision energy) using a log-derivative propagation
method of Johnson.\cite{Manolopoulos:JCP86} A total of 290 energy values in a logarithmic scale
in the range $10^{-6}-20$ cm$^{-1}$ relative to the initial $v=2,j=2$ level of D$_2$ are propagated
to compute the quenching cross sections. Total angular momentum quantum numbers $J=0-9$ are
included to secure convergence of the cross sections. Calculations were carried out on the BSP3
PES~\cite{BSP3} for the He-H$_2$ system. Additional calculations on earlier versions of the BSP PES
for the He-H$_2$ system~\cite{3D_PES} presented in the supplementary materials yielded similar
results.
\begin{figure}[b!]
\begin{center}
\includegraphics[width=1.0\linewidth]{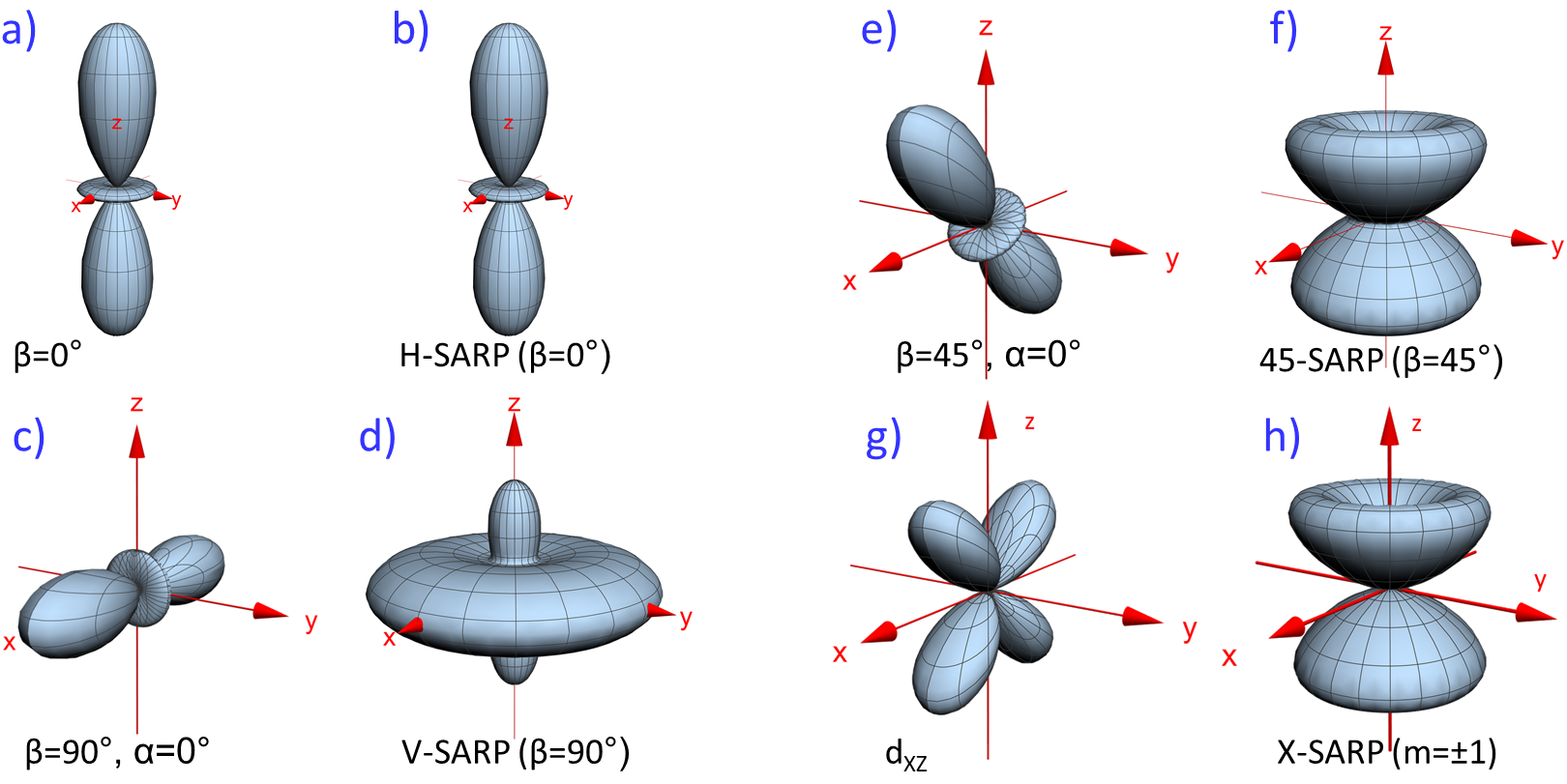}
\end{center}
\caption{Stereodynamical portraits (i.e., PDFs for the internuclear axis distribution)
for different experimental preparations. PDFs of the  1st and 3rd column depict the distributions for
given values of  $\beta$, and  $\alpha$, the angles that define the direction of the pump laser
polarization in the scattering frame (panels a, c, e). The 2nd and 4th column display the alignment distributions integrated
over $\alpha$ corresponding to the experimental PDF (panels b, d, f, and h). Note the similarities
between 45-SARP, and X-SARP preparations upon integration over the azimuthal angle. 
}
\label{fig:portrait}
\end{figure}

%

In He + D$_2$ SARP experiments, \cite{sarp_hed2,sarp_hed2_science} a  molecular beam of D$_2$ and
He is co-expanded and collimated. Then, using SARP, nearly all the D$_2$ ($v$=0,$j$=0) is pumped
into the excited D$_2$ ($v$=2,$j$=2) state. Changing the polarization direction of the pump and
Stokes laser pulse with respect to the scattering frame (defined with $z$ along the relative
velocity of the colliding partners, $\bm{k}$, and the $x-z$ plane as that containing $\bm{k}$ and
the recoil direction, $\bm{k'}$, after the collision) it is possible to produce anisotropic
distributions of the D$_2$ internuclear axis. Following Ref.~\citenum{AMHKA:JPCA05},
$P(\theta_r,\phi_r)$, the probability density function (PDF) that describes the spatial
distribution of the internuclear axis following SARP excitation is given by:
\begin{equation}\label{eq:extrinsic}
P(\theta_r,\phi_r) = \sum_{k=0}^{2j} \sum_{q=-k}^{q=k} \frac{2 k +1}{4 \pi} a^{(k)}_q
\langle j 0, k 0 | j 0 \rangle C^*_{kq} (\theta_r,\phi_r).
\end{equation}
where $\theta_r$, and $\phi_r$ are the polar and azimuthal angles that specify the direction of the
D$_2$ internuclear axis with respect to the scattering frame, $C_{kq}$ is the modified spherical
harmonic, $\langle .. | .. \rangle$ is a Clebsch-Gordan coefficient, and $a^{(k)}_q$ are the
extrinsic (preparation) polarization parameters in the $\bm k-\bm k'$ frame.  In general terms, the
initial state is prepared in the laboratory-fixed frame where $\rm Z$ is defined along the pump and
Stokes polarization vector, assuming for the time being, parallel to each other. In this frame the
polarization parameters are given by ${{A}}^{(k)}_0$ and are related to those in the scattering
frame by \cite{AMHKA:JPCA05}
\begin{equation}\label{eq:extrinsicpp}
a^{(k)}_q  = {A}^{(k)}_0  \, [D^{k}_{q 0}(\alpha, \beta, \gamma=0)]^*=  C_{kq}(\beta, \alpha) \,
{A}^{(k)}_0\,,
\end{equation}
where $\beta$ and $\alpha$ are the polar and azimuthal angles that define the direction of the
laser polarization vector ($\rm Z$ axis) in the $\bm k-\bm k'$ scattering frame. Note that because
the distributions we are considering have cylindrical symmetry around $Z$, the only non-vanishing
$A^{(k)}_Q$ moments have $Q=0$. If the prepared state is $| j \, 0\rangle$  then $A^{(k)}_0
=\langle j 0, k 0 | j 0 \rangle$. Varying the direction of the laboratory axis, $\rm Z$, with
respect to the scattering frame amounts to changing the D$_2$ alignment in the scattering
frame.\cite{AMHKA:JPCA05}

Equation \ref{eq:extrinsicpp} is only valid if the pump and Stokes pulses are parallel to each
other. This is the case of H-SARP ($\beta=0^{\circ}$), V-SARP ($\beta=90^{\circ}$), and the
uniaxial 45$^{\circ}$ SARP preparation ($\beta=45^{\circ}$). The PDFs for
H-SARP($\beta=0^{\circ}$),  $\beta=90^{\circ}$,  $\beta=45^{\circ}$, are shown in panels (a), (c),
and (e) of Fig.\ref{fig:portrait}, where the $\alpha$ angle has been chosen as zero.

In Ref.~\citenum{sarp_hed2}, Zhou et al. also prepared the so-called biaxial state (X-SARP) using a
cross-polarized pump (along $\rm Z$) and Stokes (along $\rm X$) pulses that allow to prepare a pure
state as a superposition of $|j=2, m=\pm 1 \rangle$. For the X-SARP preparation,
Eq.~\ref{eq:extrinsicpp} is not valid, and the only non-zero $ {A}^{(k)}_Q$ moments are given by
(see SI for derivation):
\begin{eqnarray}
{A}^{(0)}_0 = 1 \quad \quad {A}^{(2)}_0 = -\frac{1}{\sqrt{14}} \quad && \quad {A}^{(2)}_2 = {A}^{(2)}_{-2} =
-\frac{1}{2} \sqrt{\frac{3}{7}} \\
\nonumber {A}^{(4)}_0 = -\frac{2}{3} \sqrt{\frac{2}{7}}\quad && \quad {A}^{(4)}_2 =
{A}^{(4)}_{-2} = \frac{1}{3} \sqrt{\frac{5}{7}}\,,
\label{eq:extrinsicppm1}
\end{eqnarray}

that lead to the same PDF, $|d_{\rm XZ}|^2$, as for a d$_{\rm XZ}$ spherical harmonic in the lab
frame ($\rm{X,Y,Z}$) now defined by the directions of the pump and Stokes laser polarizations. In
the general case, to obtain the polarization parameters in the scattering frame, it would be
necessary to use the expression
\begin{equation}\label{eq:extrinsicpp2}
a^{(k)}_q  = \sum_Q [D^{k}_{q Q}(\alpha, \beta, \gamma)]^* {A}^{(k)}_Q \,.
\end{equation}
However, if the $\rm Z$ axis is made to coincide with the $z$ axis ($\equiv \bm k$) of the scattering
frame (which in the experiment is the direction of the flight axis), $\beta$=0, and the angle
$\gamma$ can be taken arbitrarily to be zero, yielding
\begin{equation}\label{eq:extrinsicpp2}
a^{(k)}_q  = \sum_Q [D^{k}_{q Q}(\alpha, 0, 0)]^* {A}^{(k)}_Q=
\sum_Q \,e^{iq\alpha} \, d^k_{q Q}(0)= {A}^{(k)}_q \,e^{iq\alpha},
\end{equation}
where $d^k_{q Q}(0)=\delta_{q Q}$  is the reduced rotation matrix for $\beta=0$. Therefore,
$a^{(k)}_q$ and ${A}^{(k)}_q$ differ only by a phase factor (the azimuthal angle $\alpha$) and
Eq.~\eqref{eq:extrinsic} for the X-SARP preparation can be written as
\begin{equation}
P(\theta_r,\phi) = \sum_{k=0}^{2j} \sum_{q=-k}^{q=k} \frac{2 k +1}{4 \pi} A^{(k)}_q \langle j
0, k 0 | j 0 \rangle C^*_{k,q} (\theta_r,\phi),
\end{equation}
where the azimuthal angle is now $\phi=\phi_r - \alpha $, and the effect of changing $\alpha$ is
the rotation around $Z$. The distribution of internuclear axis in the X-SARP case is portrayed in
panel (g) of Fig.~\ref{fig:portrait}.

In the experiments using SARP preparation \cite{sarp_hed2,sarp_hed2_science} the collision partners
were co-expanded in a single supersonic beam in an axially symmetric arrangement. Therefore,  to
properly reproduce the experimental internuclear axis distribution that corresponds to each of the
measured DCS,  integration should be carried out over the azimuthal angle, leading to the PDFs
shown in panels (b), (d), (f), and (h) of  Fig.~\ref{fig:portrait}. The integrated PDFs are, of
course, independent of the azimuthal angle.

The expression to obtain the observable DCS, i.e., the DCS for a given internuclear axis
preparation, is\cite{AMHKA:JPCA05}
\begin{equation}\label{dcsalphabeta}
{\rm d}\sigma(\theta|\beta,\alpha) = \sum_{k=0}^{2j} \sum_{q=-k}^{k} (2 k + 1 )
U^{(k)}_q(\theta) a^{(k)}_q,
\end{equation}
where the spherical tensors of rank $k$, $U^{(k)}_q(\theta)$ are the
$j$-polarization dependent differential cross sections, $j$-PDDCSs,  which in
terms of the scattering amplitudes can be written as:
\begin{equation} \label{eq:jPDDCS}
U^{(k)}_q(\theta)= \frac{1}{2j+1} \,\sum_{m',m}  f_{j'm', \,j m}(\theta)
\, f^*_{j'm', \,j m}(\theta)\, \langle j m, k q| j m + q\rangle \,,
\end{equation}
where $f_{j'm', \,j m}(\theta)$ is the scattering amplitude (and where the indices that denote the
initial and final vibrational state have been omitted for the sake of clarity). If the initial
prepared state in the laboratory frame is $|j \,\,m=0 \rangle$, {\em i.e.}, where the Stokes and
pump pulses are parallel to each other such as in H-SARP, V-SARP, and 45-SARP,
Eq.~\eqref{dcsalphabeta} can be written as:\cite{Jambrina_PCCP_2021}
\begin{equation}\label{dcsalphabeta2}
{\rm d}\sigma(\theta|\beta,\alpha) = \sum_{m'} \Big| \sum_m \,C_{j \, m}(\beta,\alpha) \,
f_{j'm', \,j m}(\theta)  \Big|^2.
\end{equation}
For the particular case of a $j$=2 $\rightarrow$ $j'$=0 transition, Eq.~\eqref{dcsalphabeta2}
reduces to
\begin{align}\label{eqj2} \nonumber
  d\sigma(\theta| \beta,\alpha) & = \left| \,\,\frac{1}{2} (3 \cos^2 \beta - 1 )   F_{0,0} -
  \Big[\sqrt{6} \sin \beta \cos \beta  \cos \alpha\, \Big] F_{0,1} \right.  \\ & \left.   +  \Big[\sqrt{\frac{3}{2}}
  \sin^2 \beta \cos 2 \alpha \Big] F_{0,2} \,\, \right|^2
\end{align}
where we have used a shorthand notation for the scattering amplitudes $f_{j'm',\, j
m}(\theta)  \equiv F_{m', m }$, and the symmetry relationship: $F_{0,m}= (-1)^{m} F_{0,-m}$.

Upon integration over the azimuthal angle, all moments with $q\ne$0 vanish, and the observable DCS
is given by:
\begin{align}\label{dcsalphabetaphi}
I_{\beta}(\theta)=&  d\sigma(\theta| \beta)/d\theta=  \sin \theta \int_0^{2 \pi}\, d\sigma(\theta|\beta, \alpha) \,{\rm d} \alpha
= \\ & \pi \sin \theta \left[ \frac{1}{2} (3 \cos^2 \beta - 1 )^2
\left|F_{0,0}\right|^2 + \nonumber \right. \\ & \left. 6 \sin^2 \beta \cos^2 \beta
  \left|F_{0,1}\right|^2 + \frac{3}{2} \sin^4 \beta \left|F_{0,2}\right|^2 \right]. \nonumber
\end{align}

Thus, the final expression for the different observable DCSs $I_{\beta}(\theta)$
are:
\begin{align}
& I_{\beta=0}(\theta) = I_{H}(\theta) = 2 \pi \sin \theta  \left|F_{0,0}\right|^2 \label{hsarp} \\
& I_{\beta=90^{\circ}}(\theta) = I_{V}(\theta) = \pi \sin \theta \left[\frac{1}{2}  \left|F_{0,0}\right|^2 + \frac{3}{2}
\left|F_{0,2}\right|^2  \right] =  \label{vsarp} \\ \nonumber & \quad \quad  \quad \quad \frac{1}{4} I_{H}(\theta) + \frac{3}{2} \pi \sin \theta  \left|F_{0,2}\right|^2  \\
& I_{\beta=45^{\circ}/135^{\circ}}(\theta) = I_{+/-}(\theta) = \pi \sin \theta \left[\frac{1}{8} \left|F_{0,0}\right|^2 + \frac{3}{2}
\left|F_{0,1}\right|^2 + \frac{3}{8} \left|F_{0,2}\right|^2 \right] \,.
 \label{xsarp}
\end{align}
For the cross-polarized X-SARP experiment, Eq.~\ref{dcsalphabeta2} is not valid, and after
integration over the azimuthal angle, Eq. \eqref{dcsalphabeta} simplifies to (see SI for further
details):
\begin{eqnarray}
I_{X}(\theta) &=& 2 \pi \sin \theta  \left|F_{0,1}\right|^2\,.
\end{eqnarray}

There are four characteristic of the observable DCS that we wish to draw our attention to:
\begin{itemize}
\item The observable DCS shows coherences between states with different $m$. These coherences
    disappear upon integration in the azimuthal angle ({\em i.e.}, terms associated with
    $q\neq$0 vanish),  making possible to compute the observable DCS as the weighted sum of the
    DCS for pure $m$ states.
\item $I_{+}(\theta)$  can be obtained as a combination of  $I_{V}(\theta)$ and $I_{X}(\theta)$
    as follows:
 \begin{equation}\label{dcs45sarpvx}
    I_{+}(\theta) = \frac{3}{4} I_{X}(\theta) + \frac{1}{4} I_{V}(\theta).
  \end{equation}

\item Assuming that the experiment integrates over the azimuthal angle, it is not possible to isolate the contribution from $m =\pm$1 without a cross-polarized experiment. If integration over azimuthal angle is not carried out, it could be possible to isolate $m =\pm$1 by setting $\beta$=54.7$^{\circ}$
    (magic angle), and $\alpha$ = 45$^{\circ}$ (or $135 ^{\circ}$).

\item As stated in Ref.~\citenum{sarp_hed2_science}, and explained in the SI, it is possible to
    decompose the wave-function associated to X-SARP ($| \psi_X \rangle$) as a superposition of
    any pair of $|\psi_{\beta} \rangle$ and  $|\psi_{\pi-\beta} \rangle$ states ($0 < \beta <
    90$). In the particular case of $\beta$=45$^{\circ}$ the resulting expression is:
\begin{equation}\label{xsarpint}
I_{X}(\theta) = \frac{2}{3} I_{+}(\theta) + \frac{1}{3} I_{\rm int}(\theta)\,,
\end{equation}
  where the interference term is given by
\begin{equation}
I_{\rm int}(\theta) = - \sin \theta \big[ \frac{\pi}{4}   \left|F_{0,0}\right|^2 -  3 \pi   \left|F_{0,1}\right|^2 +
\frac{3 \pi}{4}   \left|F_{0,2}\right|^2 \big] \,.
\end{equation}

\end{itemize}



\begin{figure*}
  \centering
  \includegraphics[width=1.0\linewidth]{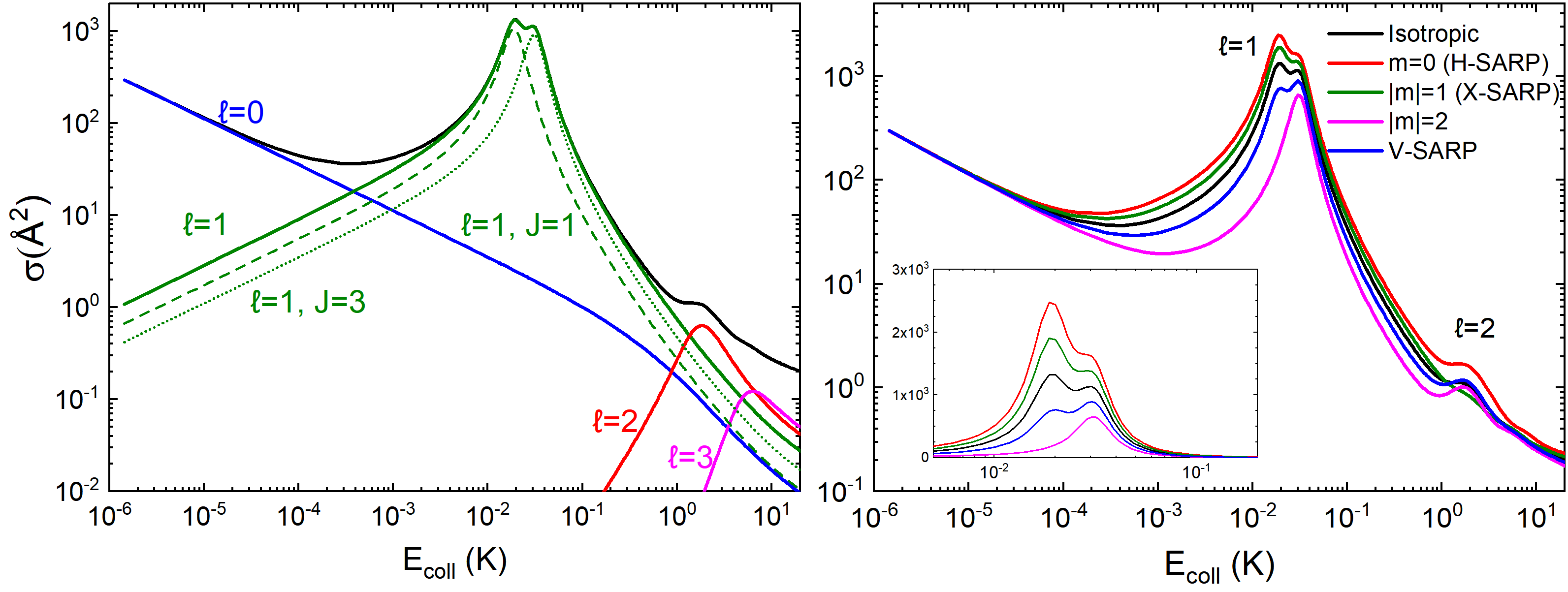}
  \caption{Left panel: Partial-wave-resolved excitation functions for He+D$_2 (v=2,j=2 \to
  v'=2,j'=0$) collisions. Right panel: Excitation functions for D$_2$ ($v=2,j=2 \to v'=2,j'=0$) by
  collisions with He for different initial preparations of the D$_2$ rotational state. The inset
  shows a zoom of the excitation function around the energy of the resonance in a linear scale.}
  \label{fig:ICS}
\end{figure*}

Before presenting the DCSs for the different experimental internuclear axis distributions, we shall
first present the unpolarized integral cross section (ICS) for the inelastic collisions between He
+ D$_2$ in the collision energy ($E_{\rm coll}$) range relevant to the experiment. The left panel
of Fig.~\ref{fig:ICS} shows the excitation functions (ICS as a function of $E_{\rm coll}$) in the
$10^{-6}-20$\, \,K  range. In agreement with previous calculations from Zhou and
Chen\cite{ICS_HD-He_Boyi} for He + D$_2$ ($v$=0,$j$=2) $\rightarrow$ He + D$_2$ ($v$=0,$j$=0) collisions on the BSP PES,\cite{3D_PES} the excitation function exhibits two salient features: i) a dominant
resonance peak at around 0.02\,\,K that results in a 30-fold increase of the cross section, and ii)
a small bump around 1.6\,\,K. Partial wave resolution of the excitation function allows us to
assign the small bump around 1.6 K to the opening of $\ell$=2 partial-wave, and the 0.02 K
resonance peak to a  $\ell$=1  resonance. Moreover, the  $\ell$=1 resonance peak is composed of two
peaks corresponding to $J$=1 ($\ell'$=1) and $J$=3 ($\ell'$=3). A similar resonance profile was observed for
other systems. \cite{HHF_Jambrina20}  Except for those features, the excitation functions at the
lowest energies (dominated by $\ell$=0, s-wave scattering) is proportional to $E_{\rm
coll}^{-1/2}$,  as expected in the Wigner threshold regime. \cite{BALAKRISHNAN19971,Lara-JCP15}

Previous studies of inelastic collisions at cold energies have shown that the intensity
of the resonance peak can be modulated by suitable alignments of the molecular bond axis.
\cite{HHF_Jambrina20,2019_PRL_Jambrina,morita_hcl-h2,Jambrina_PCCP_2021} To show
how much control can be exerted on  the $\ell$=1 resonance  by the initial preparations of
internuclear axes, the right panel of Fig. \ref{fig:ICS} displays the excitation function for  H-SARP  (in which only
$m$=0 contributes), X-SARP (where only $m=\pm$1 contribute), and V-SARP (that includes
contributions from $m$=0, and $\pm$ 2). These preparations have been depicted in
Fig.~\ref{fig:portrait}. In addition, for the sake of comparison, the excitation function for
$m$=2 is also shown.  As was demonstrated in Ref.~\citenum{Aldegunde-JCP06} there is no
control over the ICS for $\ell =$0 and, accordingly, at the Wigner regime all preparations converge
to the value of the isotropic ICS.

 For $E_{\rm coll} >$ 10$^{-4}$ K, cross sections display sensitivity to different stereodynamic preparations, and at the resonance peak H-SARP
yields larger cross sections than X-SARP and V-SARP. However,  the resonance is
prominent for the three preparations, and it is not possible to find a preparation for which the
resonance disappears, as in the case for HD + H$_2$ inelastic collisions.
\cite{2019_PRL_Jambrina} Nevertheless,  it is possible  to remove the contribution of
($\ell$=1,$J$=3)  peak, if a pure $m$=2 state could be
prepared. In the case of the smaller $\ell =$2 peak, it is the X-SARP ($m$=1) preparation that
nearly washes out the observed peak. Overall, H-SARP preparation leads to  larger  cross sections
at all energies including the $\ell$=1 and $\ell$=2 peaks, indicating that the collision
mechanisms does not change significantly with $E_{\rm coll}$, in particular at the resonance.

\begin{figure*}
\begin{center}
\includegraphics[width=1.0\linewidth]{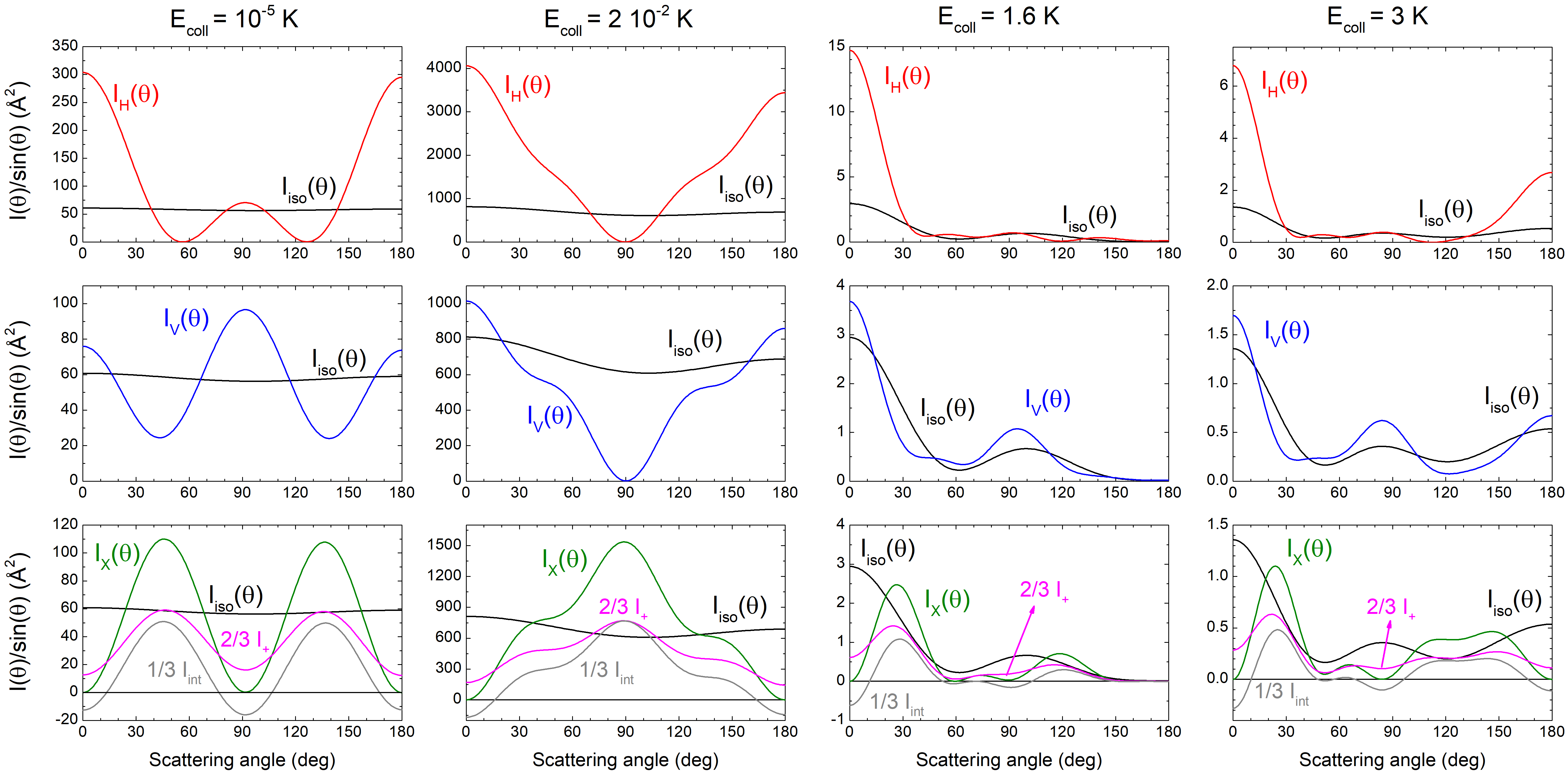}
\end{center}
\caption{Differential Cross Sections for He+D$_2$ ($v=2,j=2 \to v'=2,j'=0$) collisions
for different initial preparations of the D$_2$ rotational state at four different energies:
10$^{-5}$, $2\times 10^{-2}$, 1.6, and 3~K.  The top and middle panels show the results
integrated over the azimuthal angle for  H-SARP and V-SARP. The bottom panels show X-SARP along with  45-SARP ($I_+$) preparations, and
I$_{\rm int}$, the interference term.  In all panels the isotropic (unpolarized) DCS is shown for the sake of comparison. To see how
the shape of the DCS evolves with $E_{\rm coll}$, in Fig. S3 the normalized DCS are
 shown as a function of $E_{\rm coll}$. The latter results clearly show how the shape
 of the DCS is governed by the $\ell$=1 resonance.
}
\label{fig:DCS-energy}
\end{figure*}

To elucidate the effect of different preparations  on the DCS integrated over the
azimuthal angle, Fig.~\ref{fig:DCS-energy}  shows the differential cross sections at four
different energies: 10$^{-5}$, $2\times 10^{-2}$, 1.6, and 3~K, where  I$_{\beta} (\theta)$ are divided by $\sin \theta$ to highlight the strong  preference for  extreme forward and backward scattering (results where the  $\sin \theta$ term is retained as shown in Fig. S1).   At $E_{\rm coll}$ =10$^{-5}$ K
(well within the Wigner regime),  no control can be achieved at the integral cross section level, but
 the DCS shows sensitivity to the  different initial preparations. For the H-SARP
preparation ($m$=0) the DCS shows three salient peaks, at 0$^{\circ}$, 90$^{\circ}$, and
180$^{\circ}$, whereas the isotropic (unpolarized) DCS is essentially independent of the
scattering angle.
 The $m$=$\pm$2 component of V-SARP DCS (Eq.~\eqref{vsarp}) results in a gaussian-like function centered at 90$^{\circ}$, which when combined with the $m$=0 contribution, leads to a DCS with a prominent  90$^{\circ}$ peak.
For X-SARP, only $m$=$\pm$1 contributes, and if only $\ell =$0 is present, the DCS shows maxima at
45$^{\circ}$ and 135$^{\circ}$, and  nodes at 0$^{\circ}$, 90$^{\circ}$, and  180$^{\circ}$.
As a consequence of the connection between X-SARP, V-SARP, and 45-SARP DCSs ( Eq. \eqref{dcs45sarpvx}-\eqref{xsarpint}), the latter displays a more isotropic DCS.

As shown in Fig. \ref{fig:ICS}, at  $2\times10^{-2}$\,\,K  the scattering is dominated by the
$\ell$=1 partial wave. If only $\ell$=1 contributes, the DCS should feature a node at
90$^{\circ}$ for $m$=0, 2, and it would be forward-backward symmetric. The tiny
contribution of $\ell$=0 slightly breaks the forward-backward symmetry of the DCS, with a
non-zero value  at 90$^{\circ}$ for $m$=0,2. This minimum at 90$^{\circ}$ for H-SARP (only
$m$=0) and V-SARP ($m$=0 and 2) is then a fingerprint of the $\ell$=1 resonance. The fact
that partial-waves that barely contribute to the ICS could modulate the DCS via interference is
common, and in some cases it can even determine the shape of the DCS.
\cite{JHASJZ:NC15,JMA:CS18}  For DCS-X, there is only contribution from $|m|$=1 and, hence,
we observe a maximum at 90$^{\circ}$.

For $E_{\rm coll} > $ 1\,\,K,  $\ell$=2 begins to contribute and we enter the multiple partial wave regime with the DCS no longer
forward-backward symmetric (it is possible to obtain symmetric DCS for systems in which
many partial-waves contribute, but  an asymmetric DCS implies  more than one partial-wave).
At  $E_{\rm coll}$ = 1.6\,\,K, forward scattering is preferred, in particular for H-SARP that features a prominent forward peak. For DCS-V, $m$=2 also contributes, leading to a
noteworthy maximum at 90$^{\circ}$. For X-SARP the DCS is also predominantly forward,
although no scattering occurs at 0$^{\circ}$.  At the highest energy shown, $E_{\rm
coll}$ = 3.0\,\,K, the only change is the larger contribution of backward scattering, displaying
 a backward peak for H-SARP and V-SARP DCSs.

\begin{figure*}
\begin{center}
\includegraphics[width=1.0\linewidth]{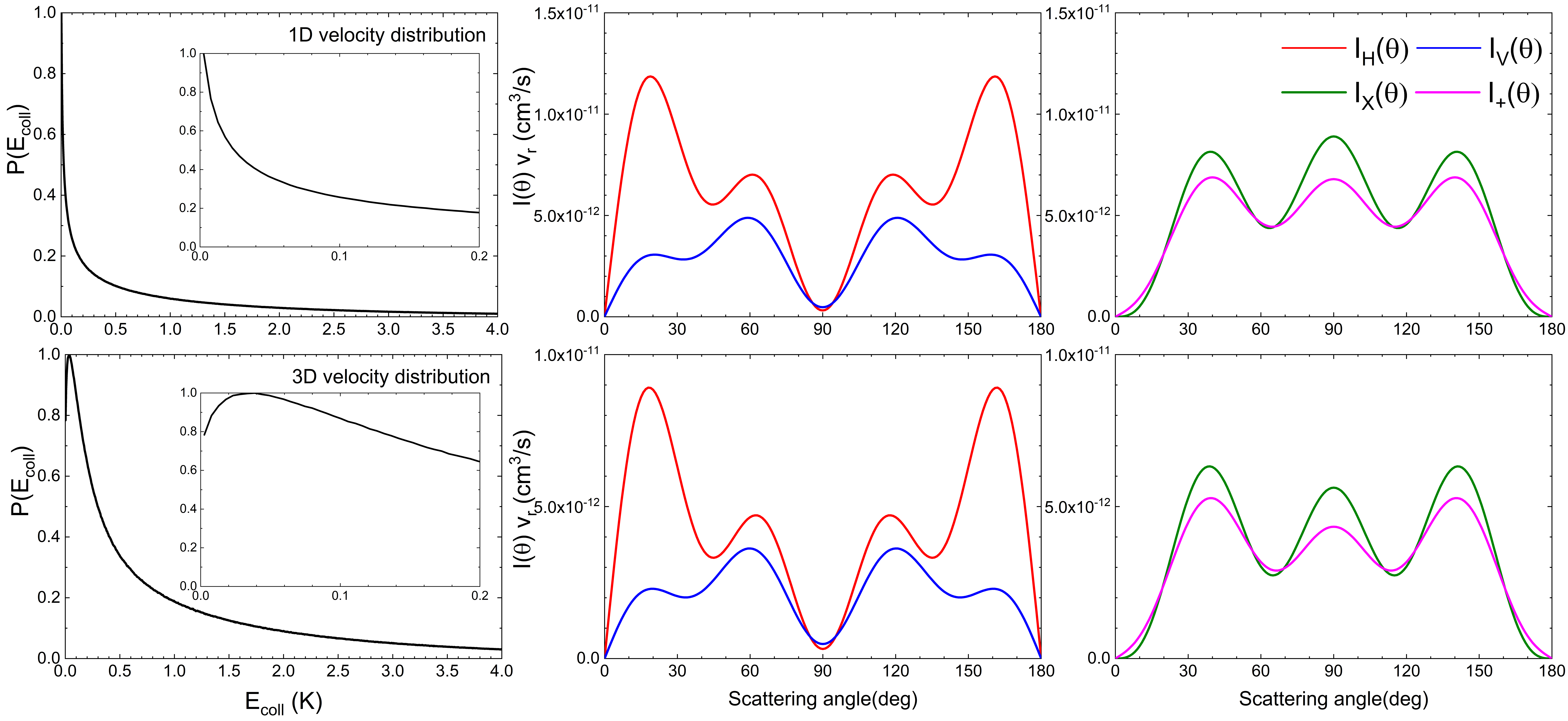}
\end{center}
\caption{Velocity-averaged differential rate coefficients for D$_2$ ($v=2,j=2 \to v'=2,j'=0$) by
collisions with He for the H-SARP,  V-SARP, X-SARP, and 45-SARP preparations of the initial
D$_2$ alignments  as a function of the scattering angle. Results are shown for two different
$E_{\rm coll}$ distributions, one assuming a 1D distribution (top panels), and another
assuming a 3D distribution (bottom panels).  The insets of the left panels display the maxima of the  $E_{\rm coll}$ distributions,
P($E_{\rm coll}$). Results assuming a 3D distribution with a larger divergence are shown in Fig. S2
}
\label{fig:DCS_aver}
\end{figure*}

As discussed earlier, the He + D$_2$ SARP experiments
\cite{sarp_hed2,sarp_hed2_science} were carried out using a single collimated molecular
beam in which D$_2$ and He were co-expanded. Based on the experimental collision speeds,
if the beam were perfectly collimated, the resulting collision energy distribution would be that shown  in the top left panel of Fig.~\ref{fig:DCS_aver} (see also the supplementary materials of Ref. \citenum{sarp_hed2}).  If  the small divergence of the beam (12 mrad=0.7$^{\circ})$ is taken into account, the resulting 3D $E_{\rm coll}$--distribution would
be that shown in the lower left panel of Fig.~\ref{fig:DCS_aver}.  In the other panels, we present the observable DCSs for different initial preparations averaged
over the  corresponding collision energy distributions. Since experimentally it is not possible
to distinguish between scattering at $\theta$ and $\pi-\theta$, the DCSs  shown here are
symmetrized as discussed in the SARP experiments~\cite{sarp_hed2,sarp_hed2_science}. Apart from the irrelevant absolute value, there is no difference in the shape of
the observable DCSs obtained for the 1D and 3D energy distribution functions. 
For both H-SARP and V-SARP the
main feature is the deep minimum observed at $\theta$=90$^{\circ}$, with four maxima at
15 and 60, 120, and 165$^{\circ}$ of different intensities. For X-SARP and 45-SARP we observe
three maxima, at 30, 90, and 150$^{\circ}$. 

\begin{figure}
\begin{center}
\includegraphics[width=1.0\linewidth]{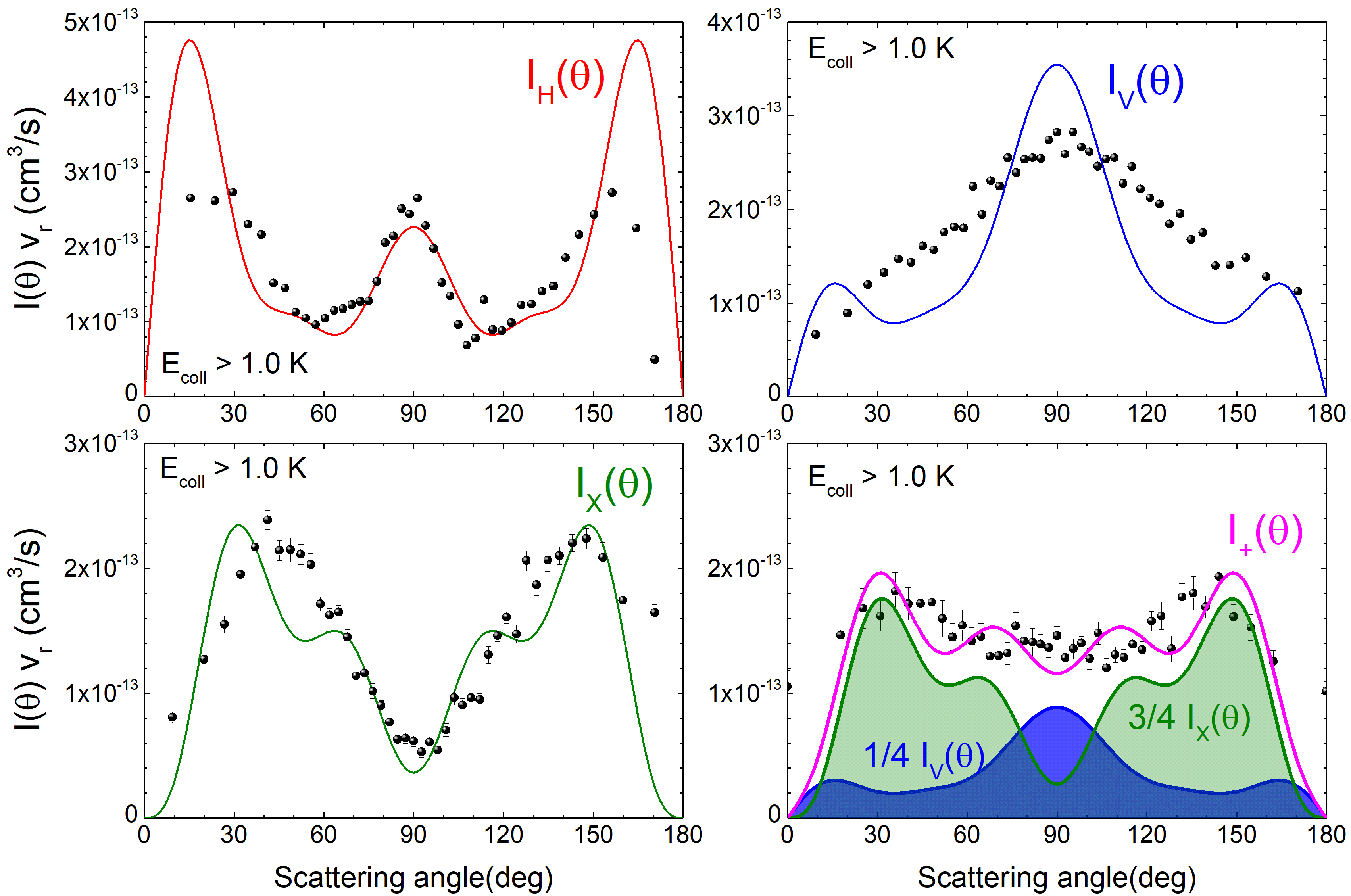}
\end{center}
\caption{Velocity-averaged differential rate coefficients for D$_2$ ($v=2,j=2 \to v'=2,j'=0$) by
collisions with He for the H-SARP,  V-SARP, X-SARP, and 45-SARP preparations of the initial
D$_2$ orientations as functions of the scattering angle. Results are shown assuming a 3D velocity
distribution, and excluding $E_{\rm coll}$ < 1 K. Experimental results from
Ref.~\citenum{sarp_hed2} and \citenum{sarp_hed2_science} are  included as black dots. As shown in Eq. \eqref{dcs45sarpvx}, $I_{+}(\theta)$ can be obtained adding 1/4 $I_{V}(\theta)$ and 3/4 $I_{X}(\theta)$, and these two
contributions are also shown in the bottom right panel.
}
\label{fig:DCS_averemin}
\end{figure}

Results presented in  Fig.~\ref{fig:DCS_aver}  are clearly at variance with the experimental results (reproduced as black dots in
Fig.~\ref{fig:DCS_averemin}). These differences are especially noteworthy at
$\theta$=90$^{\circ}$. At this angle, experimental results show maxima for H-SARP, and
V-SARP and a deep minimum for X-SARP. Our results, instead show deep minima for V-SARP,
and H-SARP and a maximum for X-SARP. The evolution of the shape of the DCS as a function of
$E_{\rm coll}$ displayed in Fig. S3 clearly indicates that these features are caused solely by the $\ell$=1 resonance. We
can artificially remove $\ell$=1 contribution, for example, excluding the averaging over
$E_{\rm coll} <$ 1\,\,K.  By doing this we obtain the results displayed in
Fig.~\ref{fig:DCS_averemin}, which reproduce the experimental results fairly well, including the effect of the interference term in X-SARP.

Altogether, our results show that there is a clear discrepancy between calculations and experiment and that this discrepancy is caused by a strong $\ell=1$ resonance whose overwhelming contribution
breaks the agreement with the experimental data from Ref.~ \citenum{sarp_hed2_science}. In
Ref.~\citenum{sarp_hed2},  based on the centrifugal energy barriers the authors estimate  that the
differential rate coefficient factor (DCS multiplied by the relative velocity)  would be roughly
10 times larger for $\ell$=2 than for $\ell$=1, which could somewhat explain why $\ell=1$
resonance is not experimentally observed. In the bottom panel of Fig. S4 we show that,
indeed,  the incoming flux at the energies of  $\ell$=2 collisions is dominant. However, the
contribution of  $\ell$=1 found in the present work is  so prevailing  that it dominates the
contribution from lower collision energies (see panels a-c of Fig. S4).

The disagreement between theory and experiment for He + D$_2$ ($v$=2,$j$=2) inelastic
collisions is unexpected. For Rg + NO($^2 \Pi$) inelastic collisions, an excellent agreement
between theory and experiment is obtained \cite{Meeraker:S20,Suits_NS22} even though
calculations for the latter system involve more electrons and potential energy surfaces, and
should be, in principle, less accurate.

To carry out our calculations we used the BSP3 PES, which is the best available
potential energy surface for He+H$_2$ collisions and it is computed with a very large-basis
set at the CCSD(T) level  augmented by corrections for higher-order excitations (up to
full configuration interaction  level) and includes the diagonal Born-Oppenheimer correction.
Thus, as far as electronic structure calculations are considered, there is a limited room for
improvement. Moreover, the BSP3 PES has been shown to yield highly accurate results for
Raman shifting and broadening cross sections for He-H$_2$ complexes in cavity
experiments~\cite{BSP3_2020_PRA}.

One could argue that the position and intensity of a resonance could be sensitive to small changes of the
interaction potential. To see if this is the case for this system, we examined the sensitivity of the $\ell$=1 peak  to the choice of the ab-initio potential used in the scattering calculations. To this end, we repeated our calculations using two additional
PESs that were developed by Garberoglio et al.~\cite{Garberoglio_HeH2} based on the original BSP
potential~\cite{3D_PES}. They are  referred to as ``BSP$+$'' and ``BSP$-$'' potentials which
are fitted to (energy+uncertainty) and  (energy-uncertainty) of the ab initio data, respectively.
These ``upper limit'' and ``lower limit'' versions of the BSP potentials were earlier used to
estimate uncertainties for the fully  quantum calculations of the second virial coefficients to
temperatures as low as 8\,\,K for  He-H$_2$ (16\,\,K for He-D$_2$ due to limited
experimental data) that yielded good  agreement with experiments~\cite{Garberoglio_HeH2}.
Using these potentials, the  position of the  $\ell$=1 resonance peak changes, in particular with the BSP- potential (See Fig. S5), but the change
is not enough to reduce the weight of $\ell$=1 collisions after averaging  over the experimental
$P(E_{\rm coll}$), as shown in Fig. S6. We  also explore the possibility of an error
in the scattering calculations. To rule out this  possibility,  we repeated the calculations for
different D$_2$ rovibrational states and also  using the ABC code~\cite{ABC_code} which uses
hyperspherical coordinates for the  scattering calculations. Again, the $\ell$=1 resonance
peak is dominant (Fig S7 and S8).


To summarize, we have carried out quantum scattering calculations for the inelastic
quenching of D$_2$ ($v$=2,$j$=2) in collisions with He atoms at cold energies. The
excitation function is governed by a sharp peak at 0.02\,\,K,  due to a
$\ell$=1 resonance. We examined if the resonance peak could be modulated changing the
relative direction of D$_2$ internuclear axis with respect to the approach direction. Although
a considerable degree of control could be achieved, it is not possible to reduce the
importance of the resonance significantly. The calculated  differential cross sections for
different internuclear axis distributions are clearly at variance with the experimental results of
Ref. \citenum{sarp_hed2,sarp_hed2_science}, unless the $\ell$=1 resonance contribution is
excluded from our calculations (by excluding contributions from
$E_{\rm coll}< 1$ K). Indeed, we obtain near-quantitative agreement with the experiment when the  $\ell$=1 resonance contribution is excluded. The disagreement between theory and experiment is
unexpected for this system due to the numerically exact nature of the quantum scattering calculations and the high-level ab-initio theory employed in the construction of the He-H$_2$ potential energy surface adopted in the calculations. Furthermore,
$\ell$=1 resonance is still dominant when the interaction potential is  modified within
its theoretical  limits, and also if other vibrational states are sampled. Further experiments and/or
calculations are needed to discern the source of this discrepancy.

\section*{Supplementary material}
See the supplementary material associated with this article for additional figures and derivation of the polarization parameters and DCS associated to X-SARP.

\begin{acknowledgments}
We are grateful to Nandini Mukherjee providing us the experimental data in Ref. \citenum{sarp_hed2,sarp_hed2_science} and for helpful
discussions. We thank Humberto da Silva Jr for useful comments. This work was supported in part by NSF grant No. PHY-2110227  (N.B.) and ARO MURI
grant No. W911NF-19-1-0283 (N.B.). P.G.J.  gratefully acknowledges grant
PID2020-113147GA-I00 funded by MCIN/AEI/10.13039/, and F.J.A. acknowledges   funding
by the Spanish Ministry of Science and Innovation (Grant No. PGC2018-096444-B-I00).
\end{acknowledgments}

\section*{Data availability statement}
The data that support the findings of this study are available within the article and its
supplementary material. Data is also available from the authors upon reasonable request.

\section*{References}

\bibliography{cite}

\begin{thebibliography}{49}%
\makeatletter
\providecommand \@ifxundefined [1]{%
 \@ifx{#1\undefined}
}%
\providecommand \@ifnum [1]{%
 \ifnum #1\expandafter \@firstoftwo
 \else \expandafter \@secondoftwo
 \fi
}%
\providecommand \@ifx [1]{%
 \ifx #1\expandafter \@firstoftwo
 \else \expandafter \@secondoftwo
 \fi
}%
\providecommand \natexlab [1]{#1}%
\providecommand \enquote  [1]{``#1''}%
\providecommand \bibnamefont  [1]{#1}%
\providecommand \bibfnamefont [1]{#1}%
\providecommand \citenamefont [1]{#1}%
\providecommand \href@noop [0]{\@secondoftwo}%
\providecommand \href [0]{\begingroup \@sanitize@url \@href}%
\providecommand \@href[1]{\@@startlink{#1}\@@href}%
\providecommand \@@href[1]{\endgroup#1\@@endlink}%
\providecommand \@sanitize@url [0]{\catcode `\\12\catcode `\$12\catcode
  `\&12\catcode `\#12\catcode `\^12\catcode `\_12\catcode `\%12\relax}%
\providecommand \@@startlink[1]{}%
\providecommand \@@endlink[0]{}%
\providecommand \url  [0]{\begingroup\@sanitize@url \@url }%
\providecommand \@url [1]{\endgroup\@href {#1}{\urlprefix }}%
\providecommand \urlprefix  [0]{URL }%
\providecommand \Eprint [0]{\href }%
\providecommand \doibase [0]{https://doi.org/}%
\providecommand \selectlanguage [0]{\@gobble}%
\providecommand \bibinfo  [0]{\@secondoftwo}%
\providecommand \bibfield  [0]{\@secondoftwo}%
\providecommand \translation [1]{[#1]}%
\providecommand \BibitemOpen [0]{}%
\providecommand \bibitemStop [0]{}%
\providecommand \bibitemNoStop [0]{.\EOS\space}%
\providecommand \EOS [0]{\spacefactor3000\relax}%
\providecommand \BibitemShut  [1]{\csname bibitem#1\endcsname}%
\let\auto@bib@innerbib\@empty
\bibitem [{\citenamefont {Wang}, \citenamefont {Lin},\ and\ \citenamefont
  {Liu}(2011)}]{Wang2011}%
  \BibitemOpen
  \bibfield  {author} {\bibinfo {author} {\bibfnamefont {F.}~\bibnamefont
  {Wang}}, \bibinfo {author} {\bibfnamefont {J.~S.}\ \bibnamefont {Lin}},\ and\
  \bibinfo {author} {\bibfnamefont {K.}~\bibnamefont {Liu}},\ }\bibfield
  {title} {\enquote {\bibinfo {title} {Steric control of the reaction of ch
  stretch--excited {CHD$_3$} with chlorine atom},}\ }\href
  {https://doi.org/10.1126/science.1199771} {\bibfield  {journal} {\bibinfo
  {journal} {Science}\ }\textbf {\bibinfo {volume} {331}},\ \bibinfo {pages}
  {900--903} (\bibinfo {year} {2011})}\BibitemShut {NoStop}%
\bibitem [{\citenamefont {Wang}, \citenamefont {Liu},\ and\ \citenamefont
  {Rakitzis}(2012)}]{Wang2012}%
  \BibitemOpen
  \bibfield  {author} {\bibinfo {author} {\bibfnamefont {F.}~\bibnamefont
  {Wang}}, \bibinfo {author} {\bibfnamefont {K.}~\bibnamefont {Liu}},\ and\
  \bibinfo {author} {\bibfnamefont {T.~P.}\ \bibnamefont {Rakitzis}},\
  }\bibfield  {title} {\enquote {\bibinfo {title} {Revealing the stereospecific
  chemistry of the reaction of cl with aligned {CHD$_3$} ($\nu$1= 1)},}\ }\href
  {https://doi.org/10.1038/nchem.1383} {\bibfield  {journal} {\bibinfo
  {journal} {Nat. Chem.}\ }\textbf {\bibinfo {volume} {4}},\ \bibinfo {pages}
  {636--641} (\bibinfo {year} {2012})}\BibitemShut {NoStop}%
\bibitem [{\citenamefont {Wang}\ and\ \citenamefont
  {Liu}(2016{\natexlab{a}})}]{Wang2016a}%
  \BibitemOpen
  \bibfield  {author} {\bibinfo {author} {\bibfnamefont {F.}~\bibnamefont
  {Wang}}\ and\ \bibinfo {author} {\bibfnamefont {K.}~\bibnamefont {Liu}},\
  }\bibfield  {title} {\enquote {\bibinfo {title} {{Differential steric effects
  in Cl reactions with aligned CHD$_3$ (v$_1$= 1) by the R (0) and Q (1)
  transitions. I. Attacking the excited C--H bond}},}\ }\href
  {https://doi.org/10.1063/1.4964652} {\bibfield  {journal} {\bibinfo
  {journal} {J. Chem. Phys.}\ }\textbf {\bibinfo {volume} {145}},\ \bibinfo
  {pages} {144305} (\bibinfo {year} {2016}{\natexlab{a}})}\BibitemShut
  {NoStop}%
\bibitem [{\citenamefont {Wang}\ and\ \citenamefont
  {Liu}(2016{\natexlab{b}})}]{Wang2016b}%
  \BibitemOpen
  \bibfield  {author} {\bibinfo {author} {\bibfnamefont {F.}~\bibnamefont
  {Wang}}\ and\ \bibinfo {author} {\bibfnamefont {K.}~\bibnamefont {Liu}},\
  }\bibfield  {title} {\enquote {\bibinfo {title} {{Differential steric effects
  in Cl reactions with aligned CHD$_3$ (v$_1$= 1) by the R (0) and Q (1)
  transitions. II. Abstracting the unexcited D-atoms}},}\ }\href
  {https://doi.org/10.1063/1.4964653} {\bibfield  {journal} {\bibinfo
  {journal} {J. Chem. Phys.}\ }\textbf {\bibinfo {volume} {145}},\ \bibinfo
  {pages} {144306} (\bibinfo {year} {2016}{\natexlab{b}})}\BibitemShut
  {NoStop}%
\bibitem [{\citenamefont {Brouard}\ \emph {et~al.}(2015)\citenamefont
  {Brouard}, \citenamefont {Chadwick}, \citenamefont {Gordon}, \citenamefont
  {Hornung}, \citenamefont {Nichols}, \citenamefont {Aoiz},\ and\ \citenamefont
  {Stolte}}]{Brouard2015}%
  \BibitemOpen
  \bibfield  {author} {\bibinfo {author} {\bibfnamefont {M.}~\bibnamefont
  {Brouard}}, \bibinfo {author} {\bibfnamefont {H.}~\bibnamefont {Chadwick}},
  \bibinfo {author} {\bibfnamefont {S.}~\bibnamefont {Gordon}}, \bibinfo
  {author} {\bibfnamefont {B.}~\bibnamefont {Hornung}}, \bibinfo {author}
  {\bibfnamefont {B.}~\bibnamefont {Nichols}}, \bibinfo {author} {\bibfnamefont
  {F.~J.}\ \bibnamefont {Aoiz}},\ and\ \bibinfo {author} {\bibfnamefont
  {S.}~\bibnamefont {Stolte}},\ }\bibfield  {title} {\enquote {\bibinfo {title}
  {Rotational orientation effects in {NO (X)+ Ar} inelastic collisions},}\
  }\href {https://doi.org/10.1021/acs.jpca.5b07846} {\bibfield  {journal}
  {\bibinfo  {journal} {J. Phys. Chem. A}\ }\textbf {\bibinfo {volume} {119}},\
  \bibinfo {pages} {12404--12416} (\bibinfo {year} {2015})}\BibitemShut
  {NoStop}%
\bibitem [{\citenamefont {Chadwick}\ \emph {et~al.}(2014)\citenamefont
  {Chadwick}, \citenamefont {Nichols}, \citenamefont {Gordon}, \citenamefont
  {Hornung}, \citenamefont {Squires}, \citenamefont {Brouard}, \citenamefont
  {K{\l}os}, \citenamefont {Alexander}, \citenamefont {Aoiz},\ and\
  \citenamefont {Stolte}}]{Chadwick2014}%
  \BibitemOpen
  \bibfield  {author} {\bibinfo {author} {\bibfnamefont {H.}~\bibnamefont
  {Chadwick}}, \bibinfo {author} {\bibfnamefont {B.}~\bibnamefont {Nichols}},
  \bibinfo {author} {\bibfnamefont {S.~D.~S.}\ \bibnamefont {Gordon}}, \bibinfo
  {author} {\bibfnamefont {B.}~\bibnamefont {Hornung}}, \bibinfo {author}
  {\bibfnamefont {E.}~\bibnamefont {Squires}}, \bibinfo {author} {\bibfnamefont
  {M.}~\bibnamefont {Brouard}}, \bibinfo {author} {\bibfnamefont
  {J.}~\bibnamefont {K{\l}os}}, \bibinfo {author} {\bibfnamefont {M.~H.}\
  \bibnamefont {Alexander}}, \bibinfo {author} {\bibfnamefont {F.~J.}\
  \bibnamefont {Aoiz}},\ and\ \bibinfo {author} {\bibfnamefont
  {S.}~\bibnamefont {Stolte}},\ }\bibfield  {title} {\enquote {\bibinfo {title}
  {{Inelastic Scattering of NO by Kr: Rotational Polarization over a
  Rainbow}},}\ }\href {https://doi.org/10.1021/jz501621c} {\bibfield  {journal}
  {\bibinfo  {journal} {J. Phys. Chem. Lett.}\ }\textbf {\bibinfo {volume}
  {5}},\ \bibinfo {pages} {3296--3301} (\bibinfo {year} {2014})}\BibitemShut
  {NoStop}%
\bibitem [{\citenamefont {Brouard}\ \emph {et~al.}(2013)\citenamefont
  {Brouard}, \citenamefont {Chadwick}, \citenamefont {Eyles}, \citenamefont
  {Hornung}, \citenamefont {Nichols}, \citenamefont {Aoiz}, \citenamefont
  {Jambrina},\ and\ \citenamefont {Stolte}}]{Brouard2013}%
  \BibitemOpen
  \bibfield  {author} {\bibinfo {author} {\bibfnamefont {M.}~\bibnamefont
  {Brouard}}, \bibinfo {author} {\bibfnamefont {H.}~\bibnamefont {Chadwick}},
  \bibinfo {author} {\bibfnamefont {C.~J.}\ \bibnamefont {Eyles}}, \bibinfo
  {author} {\bibfnamefont {B.}~\bibnamefont {Hornung}}, \bibinfo {author}
  {\bibfnamefont {B.}~\bibnamefont {Nichols}}, \bibinfo {author} {\bibfnamefont
  {F.~J.}\ \bibnamefont {Aoiz}}, \bibinfo {author} {\bibfnamefont {P.~G.}\
  \bibnamefont {Jambrina}},\ and\ \bibinfo {author} {\bibfnamefont
  {S.}~\bibnamefont {Stolte}},\ }\bibfield  {title} {\enquote {\bibinfo {title}
  {{Rotational alignment effects in NO (X)+ Ar inelastic collisions: An
  experimental study}},}\ }\href {https://doi.org/10.1063/1.4792159} {\bibfield
   {journal} {\bibinfo  {journal} {J. Chem. Phys.}\ }\textbf {\bibinfo {volume}
  {138}},\ \bibinfo {pages} {104310} (\bibinfo {year} {2013})}\BibitemShut
  {NoStop}%
\bibitem [{\citenamefont {Vogels}\ \emph {et~al.}(2018)\citenamefont {Vogels},
  \citenamefont {Karman}, \citenamefont {Klos}, \citenamefont {Besemer},
  \citenamefont {Onvlee}, \citenamefont {van~der Avoird}, \citenamefont
  {Groenenboom},\ and\ \citenamefont {van~de Meerakker}}]{VKKBOAGM:NC18}%
  \BibitemOpen
  \bibfield  {author} {\bibinfo {author} {\bibfnamefont {S.~N.}\ \bibnamefont
  {Vogels}}, \bibinfo {author} {\bibfnamefont {T.}~\bibnamefont {Karman}},
  \bibinfo {author} {\bibfnamefont {J.}~\bibnamefont {Klos}}, \bibinfo {author}
  {\bibfnamefont {M.}~\bibnamefont {Besemer}}, \bibinfo {author} {\bibfnamefont
  {J.}~\bibnamefont {Onvlee}}, \bibinfo {author} {\bibfnamefont {J.~O.}\
  \bibnamefont {van~der Avoird}}, \bibinfo {author} {\bibfnamefont {G.~C.}\
  \bibnamefont {Groenenboom}},\ and\ \bibinfo {author} {\bibfnamefont
  {S.~Y.~T.}\ \bibnamefont {van~de Meerakker}},\ }\bibfield  {title} {\enquote
  {\bibinfo {title} {{Scattering resonances in bimolecular collisions between
  NO radicals and H$_2$ challenge the theoretical gold standard}},}\ }\href
  {https://doi.org/10.1038/s41557-018-0001-3} {\bibfield  {journal} {\bibinfo
  {journal} {Nat. Chem.}\ }\textbf {\bibinfo {volume} {10}},\ \bibinfo {pages}
  {435--440} (\bibinfo {year} {2018})}\BibitemShut {NoStop}%
\bibitem [{\citenamefont {Onvlee}\ \emph {et~al.}(2017)\citenamefont {Onvlee},
  \citenamefont {Gordon}, \citenamefont {Vogels}, \citenamefont {Auth},
  \citenamefont {Karman}, \citenamefont {Nichols}, \citenamefont {van~der
  Avoird}, \citenamefont {Groenenboom}, \citenamefont {Brouard},\ and\
  \citenamefont {van~de Meerakker}}]{OGVAKNAGBM:NC17}%
  \BibitemOpen
  \bibfield  {author} {\bibinfo {author} {\bibfnamefont {J.}~\bibnamefont
  {Onvlee}}, \bibinfo {author} {\bibfnamefont {S.~D.~S.}\ \bibnamefont
  {Gordon}}, \bibinfo {author} {\bibfnamefont {S.~N.}\ \bibnamefont {Vogels}},
  \bibinfo {author} {\bibfnamefont {T.}~\bibnamefont {Auth}}, \bibinfo {author}
  {\bibfnamefont {T.}~\bibnamefont {Karman}}, \bibinfo {author} {\bibfnamefont
  {B.}~\bibnamefont {Nichols}}, \bibinfo {author} {\bibfnamefont
  {A.}~\bibnamefont {van~der Avoird}}, \bibinfo {author} {\bibfnamefont
  {G.~C.}\ \bibnamefont {Groenenboom}}, \bibinfo {author} {\bibfnamefont
  {M.}~\bibnamefont {Brouard}},\ and\ \bibinfo {author} {\bibfnamefont
  {S.~Y.~T.}\ \bibnamefont {van~de Meerakker}},\ }\bibfield  {title} {\enquote
  {\bibinfo {title} {{Imaging quantum stereodynamics through Fraunhofer
  scattering of NO radicals with rare-gas atoms}},}\ }\href
  {https://doi.org/10.1038/nchem.2640} {\bibfield  {journal} {\bibinfo
  {journal} {Nat. Chem.}\ }\textbf {\bibinfo {volume} {9}},\ \bibinfo {pages}
  {226--233} (\bibinfo {year} {2017})}\BibitemShut {NoStop}%
\bibitem [{\citenamefont {Sharples}\ \emph {et~al.}(2018)\citenamefont
  {Sharples}, \citenamefont {Leng}, \citenamefont {Luxford}, \citenamefont
  {McKendrick}, \citenamefont {Jambrina}, \citenamefont {Aoiz}, \citenamefont
  {Chandler},\ and\ \citenamefont {Costen}}]{SLLMJACC:NC18}%
  \BibitemOpen
  \bibfield  {author} {\bibinfo {author} {\bibfnamefont {T.~R.}\ \bibnamefont
  {Sharples}}, \bibinfo {author} {\bibfnamefont {J.~G.}\ \bibnamefont {Leng}},
  \bibinfo {author} {\bibfnamefont {T.~F.~M.}\ \bibnamefont {Luxford}},
  \bibinfo {author} {\bibfnamefont {K.~G.}\ \bibnamefont {McKendrick}},
  \bibinfo {author} {\bibfnamefont {P.~G.}\ \bibnamefont {Jambrina}}, \bibinfo
  {author} {\bibfnamefont {F.~J.}\ \bibnamefont {Aoiz}}, \bibinfo {author}
  {\bibfnamefont {D.~W.}\ \bibnamefont {Chandler}},\ and\ \bibinfo {author}
  {\bibfnamefont {M.~L.}\ \bibnamefont {Costen}},\ }\bibfield  {title}
  {\enquote {\bibinfo {title} {{Non-intuitive rotational reorientation in
  collisions of NO(A $^2\Sigma^+$) with Ne from direct measurement of a
  four-vector correlation}},}\ }\href
  {https://doi.org/10.1038/s41557-018-0121-9} {\bibfield  {journal} {\bibinfo
  {journal} {Nat. Chem.}\ }\textbf {\bibinfo {volume} {10}},\ \bibinfo {pages}
  {1148--1153} (\bibinfo {year} {2018})}\BibitemShut {NoStop}%
\bibitem [{\citenamefont {Perreault}, \citenamefont {Mukherjee},\ and\
  \citenamefont {Zare}(2017)}]{2017_Science_Perreault}%
  \BibitemOpen
  \bibfield  {author} {\bibinfo {author} {\bibfnamefont {W.~E.}\ \bibnamefont
  {Perreault}}, \bibinfo {author} {\bibfnamefont {N.}~\bibnamefont
  {Mukherjee}},\ and\ \bibinfo {author} {\bibfnamefont {R.~N.}\ \bibnamefont
  {Zare}},\ }\bibfield  {title} {\enquote {\bibinfo {title} {Quantum control of
  molecular collisions at 1 kelvin},}\ }\href
  {https://doi.org/10.1126/science.aao3116} {\bibfield  {journal} {\bibinfo
  {journal} {Science}\ }\textbf {\bibinfo {volume} {358}},\ \bibinfo {pages}
  {356--359} (\bibinfo {year} {2017})}\BibitemShut {NoStop}%
\bibitem [{\citenamefont {Perreault}, \citenamefont {Mukherjee},\ and\
  \citenamefont {Zare}(2018)}]{2018_NatChem_Perreault}%
  \BibitemOpen
  \bibfield  {author} {\bibinfo {author} {\bibfnamefont {W.~E.}\ \bibnamefont
  {Perreault}}, \bibinfo {author} {\bibfnamefont {N.}~\bibnamefont
  {Mukherjee}},\ and\ \bibinfo {author} {\bibfnamefont {R.~N.}\ \bibnamefont
  {Zare}},\ }\bibfield  {title} {\enquote {\bibinfo {title} {Cold
  quantum-controlled rotationally inelastic scattering of $\mathrm{HD}$ with
  $\mathrm{H}_2$ and $\mathrm{D}_2$ reveals collisional partner
  reorientation},}\ }\href {https://doi.org/10.1038/s41557-018-0028-5}
  {\bibfield  {journal} {\bibinfo  {journal} {Nat. Chem.}\ }\textbf {\bibinfo
  {volume} {10}},\ \bibinfo {pages} {561--567} (\bibinfo {year}
  {2018})}\BibitemShut {NoStop}%
\bibitem [{\citenamefont {Zhou}\ \emph
  {et~al.}(2021{\natexlab{a}})\citenamefont {Zhou}, \citenamefont {Perreault},
  \citenamefont {Mukherjee},\ and\ \citenamefont {Zare}}]{sarp_hed2}%
  \BibitemOpen
  \bibfield  {author} {\bibinfo {author} {\bibfnamefont {H.}~\bibnamefont
  {Zhou}}, \bibinfo {author} {\bibfnamefont {W.~E.}\ \bibnamefont {Perreault}},
  \bibinfo {author} {\bibfnamefont {N.}~\bibnamefont {Mukherjee}},\ and\
  \bibinfo {author} {\bibfnamefont {R.~N.}\ \bibnamefont {Zare}},\ }\bibfield
  {title} {\enquote {\bibinfo {title} {Shape resonance determined from angular
  distribution in {D}$_2$ (v=2, j=2) + {H}e $\to$ {D}$_2$ (v=2, j=0) + {H}e
  cold scattering},}\ }\href {https://doi.org/10.1063/5.0045087} {\bibfield
  {journal} {\bibinfo  {journal} {J. Chem. Phys}\ }\textbf {\bibinfo {volume}
  {154}},\ \bibinfo {pages} {104309} (\bibinfo {year}
  {2021}{\natexlab{a}})}\BibitemShut {NoStop}%
\bibitem [{\citenamefont {Heid}\ \emph {et~al.}(2019)\citenamefont {Heid},
  \citenamefont {Walpole}, \citenamefont {Brouard}, \citenamefont {Jambrina},\
  and\ \citenamefont {Aoiz}}]{HWBJA:NC19}%
  \BibitemOpen
  \bibfield  {author} {\bibinfo {author} {\bibfnamefont {C.~G.}\ \bibnamefont
  {Heid}}, \bibinfo {author} {\bibfnamefont {V.}~\bibnamefont {Walpole}},
  \bibinfo {author} {\bibfnamefont {M.}~\bibnamefont {Brouard}}, \bibinfo
  {author} {\bibfnamefont {P.~G.}\ \bibnamefont {Jambrina}},\ and\ \bibinfo
  {author} {\bibfnamefont {F.~J.}\ \bibnamefont {Aoiz}},\ }\bibfield  {title}
  {\enquote {\bibinfo {title} {{Side-impact collisions of Ar with NO}},}\
  }\href {https://doi.org/10.1038/s41557-019-0272-3} {\bibfield  {journal}
  {\bibinfo  {journal} {Nat. Chem.}\ }\textbf {\bibinfo {volume} {11}},\
  \bibinfo {pages} {662} (\bibinfo {year} {2019})}\BibitemShut {NoStop}%
\bibitem [{\citenamefont {Heid}\ \emph {et~al.}(2021)\citenamefont {Heid},
  \citenamefont {Bentham}, \citenamefont {Walpole}, \citenamefont {Jambrina},
  \citenamefont {Aoiz},\ and\ \citenamefont {Brouard}}]{CPWJAB:JPCL21}%
  \BibitemOpen
  \bibfield  {author} {\bibinfo {author} {\bibfnamefont {C.~G.}\ \bibnamefont
  {Heid}}, \bibinfo {author} {\bibfnamefont {I.~P.}\ \bibnamefont {Bentham}},
  \bibinfo {author} {\bibfnamefont {V.}~\bibnamefont {Walpole}}, \bibinfo
  {author} {\bibfnamefont {P.~G.}\ \bibnamefont {Jambrina}}, \bibinfo {author}
  {\bibfnamefont {F.~J.}\ \bibnamefont {Aoiz}},\ and\ \bibinfo {author}
  {\bibfnamefont {M.}~\bibnamefont {Brouard}},\ }\bibfield  {title} {\enquote
  {\bibinfo {title} {{Controlling the Spin-Orbit Branching Fraction in
  Molecular Collisions}},}\ }\href
  {https://doi.org/10.1021/acs.jpclett.0c02941} {\bibfield  {journal} {\bibinfo
   {journal} {J. Phys. Chem. Lett.}\ }\textbf {\bibinfo {volume} {12}},\
  \bibinfo {pages} {310} (\bibinfo {year} {2021})}\BibitemShut {NoStop}%
\bibitem [{\citenamefont {Heid}\ \emph {et~al.}(2020)\citenamefont {Heid},
  \citenamefont {Bentham}, \citenamefont {Walpole}, \citenamefont {Gheorge},
  \citenamefont {Jambrina}, \citenamefont {Aoiz},\ and\ \citenamefont
  {Brouard}}]{HBWGJAB:PCCP20}%
  \BibitemOpen
  \bibfield  {author} {\bibinfo {author} {\bibfnamefont {C.~G.}\ \bibnamefont
  {Heid}}, \bibinfo {author} {\bibfnamefont {I.~P.}\ \bibnamefont {Bentham}},
  \bibinfo {author} {\bibfnamefont {V.}~\bibnamefont {Walpole}}, \bibinfo
  {author} {\bibfnamefont {R.}~\bibnamefont {Gheorge}}, \bibinfo {author}
  {\bibfnamefont {P.~G.}\ \bibnamefont {Jambrina}}, \bibinfo {author}
  {\bibfnamefont {F.~J.}\ \bibnamefont {Aoiz}},\ and\ \bibinfo {author}
  {\bibfnamefont {M.}~\bibnamefont {Brouard}},\ }\bibfield  {title} {\enquote
  {\bibinfo {title} {{Probing the location of the unpaired electron in
  spin-orbit changing collisions of NO with Ar}},}\ }\href
  {https://doi.org/10.1039/D0CP04228E} {\bibfield  {journal} {\bibinfo
  {journal} {Phys. Chem. Chem. Phys.}\ }\textbf {\bibinfo {volume} {22}},\
  \bibinfo {pages} {22289} (\bibinfo {year} {2020})}\BibitemShut {NoStop}%
\bibitem [{\citenamefont {Walpole}\ \emph {et~al.}(2019)\citenamefont
  {Walpole}, \citenamefont {Heid}, \citenamefont {Jambrina}, \citenamefont
  {Aoiz},\ and\ \citenamefont {Brouard}}]{WHJAB:JPCA19}%
  \BibitemOpen
  \bibfield  {author} {\bibinfo {author} {\bibfnamefont {V.}~\bibnamefont
  {Walpole}}, \bibinfo {author} {\bibfnamefont {C.~G.}\ \bibnamefont {Heid}},
  \bibinfo {author} {\bibfnamefont {P.~G.}\ \bibnamefont {Jambrina}}, \bibinfo
  {author} {\bibfnamefont {F.~J.}\ \bibnamefont {Aoiz}},\ and\ \bibinfo
  {author} {\bibfnamefont {M.}~\bibnamefont {Brouard}},\ }\bibfield  {title}
  {\enquote {\bibinfo {title} {{Steric Effects in the Inelastic Scattering of
  NO (X)+ Ar: Side-on Orientation}},}\ }\href
  {https://doi.org/10.1021/acs.jpca.9b07264} {\bibfield  {journal} {\bibinfo
  {journal} {J. Phys. Chem. A}\ }\textbf {\bibinfo {volume} {123}},\ \bibinfo
  {pages} {8787} (\bibinfo {year} {2019})}\BibitemShut {NoStop}%
\bibitem [{\citenamefont {Zhou}\ \emph
  {et~al.}(2021{\natexlab{b}})\citenamefont {Zhou}, \citenamefont {Perreault},
  \citenamefont {Mukherjee},\ and\ \citenamefont {Zare}}]{sarp_hed2_science}%
  \BibitemOpen
  \bibfield  {author} {\bibinfo {author} {\bibfnamefont {H.}~\bibnamefont
  {Zhou}}, \bibinfo {author} {\bibfnamefont {W.~E.}\ \bibnamefont {Perreault}},
  \bibinfo {author} {\bibfnamefont {N.}~\bibnamefont {Mukherjee}},\ and\
  \bibinfo {author} {\bibfnamefont {R.~N.}\ \bibnamefont {Zare}},\ }\bibfield
  {title} {\enquote {\bibinfo {title} {Quantum mechanical double slit for
  molecular scattering},}\ }\href
  {https://www.science.org/doi/10.1126/science.abl4143} {\bibfield  {journal}
  {\bibinfo  {journal} {Science}\ }\textbf {\bibinfo {volume} {374}},\ \bibinfo
  {pages} {960--964} (\bibinfo {year} {2021}{\natexlab{b}})}\BibitemShut
  {NoStop}%
\bibitem [{\citenamefont {Perreault}, \citenamefont {Mukherjee},\ and\
  \citenamefont {Zare}(2019)}]{SARP_HD-He}%
  \BibitemOpen
  \bibfield  {author} {\bibinfo {author} {\bibfnamefont {W.~E.}\ \bibnamefont
  {Perreault}}, \bibinfo {author} {\bibfnamefont {N.}~\bibnamefont
  {Mukherjee}},\ and\ \bibinfo {author} {\bibfnamefont {R.~N.}\ \bibnamefont
  {Zare}},\ }\bibfield  {title} {\enquote {\bibinfo {title} {$\mathrm{HD}$ ($v
  = 1$, $j = 2$, $m$) orientation controls $\mathrm{HD}$-$\mathrm{He}$
  rotationally inelastic scattering near 1 $\mathrm{K}$},}\ }\href
  {https://doi.org/10.1063/1.5096531} {\bibfield  {journal} {\bibinfo
  {journal} {J. Chem. Phys.}\ }\textbf {\bibinfo {volume} {150}},\ \bibinfo
  {pages} {174301} (\bibinfo {year} {2019})}\BibitemShut {NoStop}%
\bibitem [{\citenamefont {Perreault}\ \emph {et~al.}(2021)\citenamefont
  {Perreault}, \citenamefont {Zhou}, \citenamefont {Mukherjee},\ and\
  \citenamefont {Zare}}]{Frontiers21Zare}%
  \BibitemOpen
  \bibfield  {author} {\bibinfo {author} {\bibfnamefont {W.~E.}\ \bibnamefont
  {Perreault}}, \bibinfo {author} {\bibfnamefont {H.}~\bibnamefont {Zhou}},
  \bibinfo {author} {\bibfnamefont {N.}~\bibnamefont {Mukherjee}},\ and\
  \bibinfo {author} {\bibfnamefont {R.~N.}\ \bibnamefont {Zare}},\ }\bibfield
  {title} {\enquote {\bibinfo {title} {{A Bi-Axial Quantum State That Controls
  Molecular Collisions Like a Double-Slit Interferometer}},}\ }\href
  {https://www.frontiersin.org/article/10.3389/fphy.2021.671997} {\bibfield
  {journal} {\bibinfo  {journal} {Frontiers in Physics}\ }\textbf {\bibinfo
  {volume} {9}} (\bibinfo {year} {2021})}\BibitemShut {NoStop}%
\bibitem [{\citenamefont {Vogels}\ \emph {et~al.}(2015)\citenamefont {Vogels},
  \citenamefont {Onvlee}, \citenamefont {Chefdeville}, \citenamefont {van~der
  Avoird}, \citenamefont {Groenenboom},\ and\ \citenamefont {van~de
  Meerakker}}]{Meeraker:S15}%
  \BibitemOpen
  \bibfield  {author} {\bibinfo {author} {\bibfnamefont {S.~N.}\ \bibnamefont
  {Vogels}}, \bibinfo {author} {\bibfnamefont {J.}~\bibnamefont {Onvlee}},
  \bibinfo {author} {\bibfnamefont {S.}~\bibnamefont {Chefdeville}}, \bibinfo
  {author} {\bibfnamefont {A.}~\bibnamefont {van~der Avoird}}, \bibinfo
  {author} {\bibfnamefont {G.~C.}\ \bibnamefont {Groenenboom}},\ and\ \bibinfo
  {author} {\bibfnamefont {S.~Y.~T.}\ \bibnamefont {van~de Meerakker}},\
  }\bibfield  {title} {\enquote {\bibinfo {title} {{Imaging resonances in
  low-energy NO-He inelastic collisions}},}\ }\href
  {https://doi.org/10.1126/science.aad2356} {\bibfield  {journal} {\bibinfo
  {journal} {Science}\ }\textbf {\bibinfo {volume} {350}},\ \bibinfo {pages}
  {787--790} (\bibinfo {year} {2015})}\BibitemShut {NoStop}%
\bibitem [{\citenamefont {de~Jongh}\ \emph {et~al.}(2020)\citenamefont
  {de~Jongh}, \citenamefont {Besemer}, \citenamefont {Shuai}, \citenamefont
  {Karman}, \citenamefont {van~der Avoird}, \citenamefont {Groenenboom},\ and\
  \citenamefont {van~de Meerakker}}]{Meeraker:S20}%
  \BibitemOpen
  \bibfield  {author} {\bibinfo {author} {\bibfnamefont {T.}~\bibnamefont
  {de~Jongh}}, \bibinfo {author} {\bibfnamefont {M.}~\bibnamefont {Besemer}},
  \bibinfo {author} {\bibfnamefont {Q.}~\bibnamefont {Shuai}}, \bibinfo
  {author} {\bibfnamefont {T.}~\bibnamefont {Karman}}, \bibinfo {author}
  {\bibfnamefont {A.}~\bibnamefont {van~der Avoird}}, \bibinfo {author}
  {\bibfnamefont {G.~C.}\ \bibnamefont {Groenenboom}},\ and\ \bibinfo {author}
  {\bibfnamefont {S.~Y.~T.}\ \bibnamefont {van~de Meerakker}},\ }\bibfield
  {title} {\enquote {\bibinfo {title} {{Imaging the onset of the resonance
  regime in low-energy NO-He collisions}},}\ }\href
  {https://doi.org/10.1126/science.aba3990} {\bibfield  {journal} {\bibinfo
  {journal} {Science}\ }\textbf {\bibinfo {volume} {368}},\ \bibinfo {pages}
  {626--630} (\bibinfo {year} {2020})}\BibitemShut {NoStop}%
\bibitem [{\citenamefont {Croft}\ \emph {et~al.}(2018)\citenamefont {Croft},
  \citenamefont {Balakrishnan}, \citenamefont {Huang},\ and\ \citenamefont
  {Guo}}]{2018_PRL_Croft}%
  \BibitemOpen
  \bibfield  {author} {\bibinfo {author} {\bibfnamefont {J.~F.~E.}\
  \bibnamefont {Croft}}, \bibinfo {author} {\bibfnamefont {N.}~\bibnamefont
  {Balakrishnan}}, \bibinfo {author} {\bibfnamefont {M.}~\bibnamefont
  {Huang}},\ and\ \bibinfo {author} {\bibfnamefont {H.}~\bibnamefont {Guo}},\
  }\bibfield  {title} {\enquote {\bibinfo {title} {Unraveling the
  stereodynamics of cold controlled $\mathrm{HD}$-$\mathrm{H}_{2}$
  collisions},}\ }\href {https://doi.org/10.1103/PhysRevLett.121.113401}
  {\bibfield  {journal} {\bibinfo  {journal} {Phys. Rev. Lett.}\ }\textbf
  {\bibinfo {volume} {121}},\ \bibinfo {pages} {113401} (\bibinfo {year}
  {2018})}\BibitemShut {NoStop}%
\bibitem [{\citenamefont {Croft}\ and\ \citenamefont
  {Balakrishnan}(2019)}]{2019_JCP_Croft}%
  \BibitemOpen
  \bibfield  {author} {\bibinfo {author} {\bibfnamefont {J.~F.~E.}\
  \bibnamefont {Croft}}\ and\ \bibinfo {author} {\bibfnamefont
  {N.}~\bibnamefont {Balakrishnan}},\ }\bibfield  {title} {\enquote {\bibinfo
  {title} {Controlling rotational quenching rates in cold molecular
  collisions},}\ }\href {https://doi.org/10.1063/1.5091576} {\bibfield
  {journal} {\bibinfo  {journal} {J. Chem. Phys.}\ }\textbf {\bibinfo {volume}
  {150}},\ \bibinfo {pages} {164302} (\bibinfo {year} {2019})}\BibitemShut
  {NoStop}%
\bibitem [{\citenamefont {Jambrina}\ \emph {et~al.}(2019)\citenamefont
  {Jambrina}, \citenamefont {Croft}, \citenamefont {Guo}, \citenamefont
  {Brouard}, \citenamefont {Balakrishnan},\ and\ \citenamefont
  {Aoiz}}]{2019_PRL_Jambrina}%
  \BibitemOpen
  \bibfield  {author} {\bibinfo {author} {\bibfnamefont {P.~G.}\ \bibnamefont
  {Jambrina}}, \bibinfo {author} {\bibfnamefont {J.~F.~E.}\ \bibnamefont
  {Croft}}, \bibinfo {author} {\bibfnamefont {H.}~\bibnamefont {Guo}}, \bibinfo
  {author} {\bibfnamefont {M.}~\bibnamefont {Brouard}}, \bibinfo {author}
  {\bibfnamefont {N.}~\bibnamefont {Balakrishnan}},\ and\ \bibinfo {author}
  {\bibfnamefont {F.~J.}\ \bibnamefont {Aoiz}},\ }\bibfield  {title} {\enquote
  {\bibinfo {title} {Stereodynamical control of a quantum scattering resonance
  in cold molecular collisions},}\ }\href
  {https://doi.org/10.1103/PhysRevLett.123.043401} {\bibfield  {journal}
  {\bibinfo  {journal} {Phys. Rev. Lett.}\ }\textbf {\bibinfo {volume} {123}},\
  \bibinfo {pages} {043401} (\bibinfo {year} {2019})}\BibitemShut {NoStop}%
\bibitem [{\citenamefont {Morita}\ and\ \citenamefont
  {Balakrishnan}(2020)}]{Morita_He-HD}%
  \BibitemOpen
  \bibfield  {author} {\bibinfo {author} {\bibfnamefont {M.}~\bibnamefont
  {Morita}}\ and\ \bibinfo {author} {\bibfnamefont {N.}~\bibnamefont
  {Balakrishnan}},\ }\bibfield  {title} {\enquote {\bibinfo {title}
  {Stereodynamics of rotationally inelastic scattering in cold
  $\mathrm{He}$+$\mathrm{HD}$ collisions},}\ }\href
  {https://doi.org/10.1063/5.0022190} {\bibfield  {journal} {\bibinfo
  {journal} {J. Chem. Phys.}\ }\textbf {\bibinfo {volume} {153}},\ \bibinfo
  {pages} {091101} (\bibinfo {year} {2020})}\BibitemShut {NoStop}%
\bibitem [{\citenamefont {Morita}\ \emph {et~al.}(2020)\citenamefont {Morita},
  \citenamefont {Yao}, \citenamefont {Xie}, \citenamefont {Guo},\ and\
  \citenamefont {Balakrishnan}}]{morita_hcl-h2}%
  \BibitemOpen
  \bibfield  {author} {\bibinfo {author} {\bibfnamefont {M.}~\bibnamefont
  {Morita}}, \bibinfo {author} {\bibfnamefont {Q.}~\bibnamefont {Yao}},
  \bibinfo {author} {\bibfnamefont {C.}~\bibnamefont {Xie}}, \bibinfo {author}
  {\bibfnamefont {H.}~\bibnamefont {Guo}},\ and\ \bibinfo {author}
  {\bibfnamefont {N.}~\bibnamefont {Balakrishnan}},\ }\bibfield  {title}
  {\enquote {\bibinfo {title} {Stereodynamic control of overlapping resonances
  in cold molecular collisions},}\ }\href
  {https://doi.org/10.1103/PhysRevResearch.2.032018} {\bibfield  {journal}
  {\bibinfo  {journal} {Phys. Rev. Research}\ }\textbf {\bibinfo {volume}
  {2}},\ \bibinfo {pages} {032018} (\bibinfo {year} {2020})}\BibitemShut
  {NoStop}%
\bibitem [{\citenamefont {Jambrina}\ \emph {et~al.}(2021)\citenamefont
  {Jambrina}, \citenamefont {Croft}, \citenamefont {Balakrishnan},\ and\
  \citenamefont {Aoiz}}]{Jambrina_PCCP_2021}%
  \BibitemOpen
  \bibfield  {author} {\bibinfo {author} {\bibfnamefont {P.~G.}\ \bibnamefont
  {Jambrina}}, \bibinfo {author} {\bibfnamefont {J.~F.~E.}\ \bibnamefont
  {Croft}}, \bibinfo {author} {\bibfnamefont {N.}~\bibnamefont
  {Balakrishnan}},\ and\ \bibinfo {author} {\bibfnamefont {F.~J.}\ \bibnamefont
  {Aoiz}},\ }\bibfield  {title} {\enquote {\bibinfo {title} {Stereodynamic
  control of cold rotationally inelastic co + hd collisions},}\ }\href
  {https://doi.org/10.1039/D1CP02755G} {\bibfield  {journal} {\bibinfo
  {journal} {Phys. Chem. Chem. Phys.}\ }\textbf {\bibinfo {volume} {23}},\
  \bibinfo {pages} {19364} (\bibinfo {year} {2021})}\BibitemShut {NoStop}%
\bibitem [{\citenamefont {Thibault}\ \emph {et~al.}(2017)\citenamefont
  {Thibault}, \citenamefont {Patkowski}, \citenamefont {\.{Z}uchowski},
  \citenamefont {J\'{o}\'{z}wiak}, \citenamefont {Ciury\l{}o},\ and\
  \citenamefont {Wcis\l{}o}}]{BSP3}%
  \BibitemOpen
  \bibfield  {author} {\bibinfo {author} {\bibfnamefont {F.}~\bibnamefont
  {Thibault}}, \bibinfo {author} {\bibfnamefont {K.}~\bibnamefont {Patkowski}},
  \bibinfo {author} {\bibfnamefont {P.~S.}\ \bibnamefont {\.{Z}uchowski}},
  \bibinfo {author} {\bibfnamefont {H.}~\bibnamefont {J\'{o}\'{z}wiak}},
  \bibinfo {author} {\bibfnamefont {R.}~\bibnamefont {Ciury\l{}o}},\ and\
  \bibinfo {author} {\bibfnamefont {P.}~\bibnamefont {Wcis\l{}o}},\ }\bibfield
  {title} {\enquote {\bibinfo {title} {Rovibrational line-shape parameters for
  $\mathrm{H}_2$ in $\mathrm{He}$ and new $\mathrm{H}_2$-$\mathrm{He}$
  potential energy surface},}\ }\href
  {https://doi.org/https://doi.org/10.1016/j.jqsrt.2017.08.014} {\bibfield
  {journal} {\bibinfo  {journal} {J. Quant. Spectrosc. Radiat. Transf.}\
  }\textbf {\bibinfo {volume} {202}},\ \bibinfo {pages} {308} (\bibinfo {year}
  {2017})}\BibitemShut {NoStop}%
\bibitem [{\citenamefont {S\l{}owi\ifmmode~\acute{n}\else \'{n}\fi{}ski}\ \emph
  {et~al.}(2020)\citenamefont {S\l{}owi\ifmmode~\acute{n}\else \'{n}\fi{}ski},
  \citenamefont {Thibault}, \citenamefont {Tan}, \citenamefont {Wang},
  \citenamefont {Liu}, \citenamefont {Hu}, \citenamefont {Kassi}, \citenamefont
  {Campargue}, \citenamefont {Konefa\l{}}, \citenamefont
  {J\'o\ifmmode~\acute{z}\else \'{z}\fi{}wiak}, \citenamefont {Patkowski},
  \citenamefont {\ifmmode~\dot{Z}\else \.{Z}\fi{}uchowski}, \citenamefont
  {Ciury\l{}o}, \citenamefont {Lisak},\ and\ \citenamefont
  {Wcis\l{}o}}]{BSP3_2020_PRA}%
  \BibitemOpen
  \bibfield  {author} {\bibinfo {author} {\bibfnamefont {M.}~\bibnamefont
  {S\l{}owi\ifmmode~\acute{n}\else \'{n}\fi{}ski}}, \bibinfo {author}
  {\bibfnamefont {F.}~\bibnamefont {Thibault}}, \bibinfo {author}
  {\bibfnamefont {Y.}~\bibnamefont {Tan}}, \bibinfo {author} {\bibfnamefont
  {J.}~\bibnamefont {Wang}}, \bibinfo {author} {\bibfnamefont {A.-W.}\
  \bibnamefont {Liu}}, \bibinfo {author} {\bibfnamefont {S.-M.}\ \bibnamefont
  {Hu}}, \bibinfo {author} {\bibfnamefont {S.}~\bibnamefont {Kassi}}, \bibinfo
  {author} {\bibfnamefont {A.}~\bibnamefont {Campargue}}, \bibinfo {author}
  {\bibfnamefont {M.}~\bibnamefont {Konefa\l{}}}, \bibinfo {author}
  {\bibfnamefont {H.}~\bibnamefont {J\'o\ifmmode~\acute{z}\else
  \'{z}\fi{}wiak}}, \bibinfo {author} {\bibfnamefont {K.}~\bibnamefont
  {Patkowski}}, \bibinfo {author} {\bibfnamefont {P.}~\bibnamefont
  {\ifmmode~\dot{Z}\else \.{Z}\fi{}uchowski}}, \bibinfo {author} {\bibfnamefont
  {R.}~\bibnamefont {Ciury\l{}o}}, \bibinfo {author} {\bibfnamefont
  {D.}~\bibnamefont {Lisak}},\ and\ \bibinfo {author} {\bibfnamefont
  {P.}~\bibnamefont {Wcis\l{}o}},\ }\bibfield  {title} {\enquote {\bibinfo
  {title} {$\mathrm{H}_2$-$\mathrm{He}$ collisions: Ab initio theory meets
  cavity-enhanced spectra},}\ }\href
  {https://doi.org/10.1103/PhysRevA.101.052705} {\bibfield  {journal} {\bibinfo
   {journal} {Phys. Rev. A}\ }\textbf {\bibinfo {volume} {101}},\ \bibinfo
  {pages} {052705} (\bibinfo {year} {2020})}\BibitemShut {NoStop}%
\bibitem [{\citenamefont {Martínez}\ \emph {et~al.}(2018)\citenamefont
  {Martínez}, \citenamefont {Bermejo}, \citenamefont {Thibault},\ and\
  \citenamefont {Wcisło}}]{Martinez2018}%
  \BibitemOpen
  \bibfield  {author} {\bibinfo {author} {\bibfnamefont {R.}~\bibnamefont
  {Martínez}}, \bibinfo {author} {\bibfnamefont {D.}~\bibnamefont {Bermejo}},
  \bibinfo {author} {\bibfnamefont {F.}~\bibnamefont {Thibault}},\ and\
  \bibinfo {author} {\bibfnamefont {P.}~\bibnamefont {Wcisło}},\ }\bibfield
  {title} {\enquote {\bibinfo {title} {Testing the ab initio quantum-scattering
  calculations for the d2–he benchmark system with stimulated raman
  spectroscopy},}\ }\href {https://doi.org/10.1002/jrs.5391} {\bibfield
  {journal} {\bibinfo  {journal} {J. Raman Spectrosc}\ }\textbf {\bibinfo
  {volume} {49}},\ \bibinfo {pages} {1339--1349} (\bibinfo {year} {2018})},\
  \bibinfo {note} {cited By 20}\BibitemShut {NoStop}%
\bibitem [{\citenamefont {Blatt}\ and\ \citenamefont
  {Biedenharn}(1952)}]{1952_Blatt_Biedenharn_Rev}%
  \BibitemOpen
  \bibfield  {author} {\bibinfo {author} {\bibfnamefont {J.~M.}\ \bibnamefont
  {Blatt}}\ and\ \bibinfo {author} {\bibfnamefont {L.~C.}\ \bibnamefont
  {Biedenharn}},\ }\bibfield  {title} {\enquote {\bibinfo {title} {The angular
  distribution of scattering and reaction cross sections},}\ }\href
  {https://doi.org/10.1103/RevModPhys.24.258} {\bibfield  {journal} {\bibinfo
  {journal} {Rev. Mod. Phys.}\ }\textbf {\bibinfo {volume} {24}},\ \bibinfo
  {pages} {258--272} (\bibinfo {year} {1952})}\BibitemShut {NoStop}%
\bibitem [{\citenamefont {Arthurs}\ and\ \citenamefont
  {Dalgarno}(1960)}]{1960_Dalgarno_Arthurs}%
  \BibitemOpen
  \bibfield  {author} {\bibinfo {author} {\bibfnamefont {A.~M.}\ \bibnamefont
  {Arthurs}}\ and\ \bibinfo {author} {\bibfnamefont {A.}~\bibnamefont
  {Dalgarno}},\ }\bibfield  {title} {\enquote {\bibinfo {title} {The theory of
  scattering by a rigid rotator},}\ }\href
  {https://doi.org/10.1098/rspa.1960.0125} {\bibfield  {journal} {\bibinfo
  {journal} {Proc. R. Soc. London A}\ }\textbf {\bibinfo {volume} {256}},\
  \bibinfo {pages} {540--551} (\bibinfo {year} {1960})}\BibitemShut {NoStop}%
\bibitem [{\citenamefont {Alexander}, \citenamefont {Dagdigian},\ and\
  \citenamefont {DePristo}(1977)}]{1977_Alexander_1}%
  \BibitemOpen
  \bibfield  {author} {\bibinfo {author} {\bibfnamefont {M.~H.}\ \bibnamefont
  {Alexander}}, \bibinfo {author} {\bibfnamefont {P.~J.}\ \bibnamefont
  {Dagdigian}},\ and\ \bibinfo {author} {\bibfnamefont {A.~E.}\ \bibnamefont
  {DePristo}},\ }\bibfield  {title} {\enquote {\bibinfo {title} {Quantum
  interpretation of fully state-selected rotationally inelastic collision
  experiments},}\ }\href {https://doi.org/10.1063/1.433611} {\bibfield
  {journal} {\bibinfo  {journal} {J. Chem. Phys.}\ }\textbf {\bibinfo {volume}
  {66}},\ \bibinfo {pages} {59--66} (\bibinfo {year} {1977})}\BibitemShut
  {NoStop}%
\bibitem [{\citenamefont {Alexander}(1977)}]{1977_Alexander_2}%
  \BibitemOpen
  \bibfield  {author} {\bibinfo {author} {\bibfnamefont {M.~H.}\ \bibnamefont
  {Alexander}},\ }\bibfield  {title} {\enquote {\bibinfo {title}
  {Close-coupling studies of the orientation dependence of rotationally
  inelastic collisions},}\ }\href {https://doi.org/10.1063/1.435184} {\bibfield
   {journal} {\bibinfo  {journal} {J. Chem. Phys.}\ }\textbf {\bibinfo {volume}
  {67}},\ \bibinfo {pages} {2703--2712} (\bibinfo {year} {1977})}\BibitemShut
  {NoStop}%
\bibitem [{\citenamefont {Hutson}\ and\ \citenamefont {Green}(1994)}]{MOLSCAT}%
  \BibitemOpen
  \bibfield  {author} {\bibinfo {author} {\bibfnamefont {J.~M.}\ \bibnamefont
  {Hutson}}\ and\ \bibinfo {author} {\bibfnamefont {S.}~\bibnamefont {Green}},\
  }\href@noop {} {\emph {\bibinfo {title} {MOLSCAT v.14}}},\ \bibinfo
  {organization} {Swindon: Engineering and Physical Sciences Research Council}
  (\bibinfo {year} {1994})\BibitemShut {NoStop}%
\bibitem [{\citenamefont {Skouteris}, \citenamefont {Castillo},\ and\
  \citenamefont {Manolopoulos}(2000)}]{ABC_code}%
  \BibitemOpen
  \bibfield  {author} {\bibinfo {author} {\bibfnamefont {D.}~\bibnamefont
  {Skouteris}}, \bibinfo {author} {\bibfnamefont {J.}~\bibnamefont
  {Castillo}},\ and\ \bibinfo {author} {\bibfnamefont {D.}~\bibnamefont
  {Manolopoulos}},\ }\bibfield  {title} {\enquote {\bibinfo {title} {Abc: a
  quantum reactive scattering program},}\ }\href
  {https://doi.org/https://doi.org/10.1016/S0010-4655(00)00167-3} {\bibfield
  {journal} {\bibinfo  {journal} {Comp. Phys. Comm.}\ }\textbf {\bibinfo
  {volume} {133}},\ \bibinfo {pages} {128--135} (\bibinfo {year}
  {2000})}\BibitemShut {NoStop}%
\bibitem [{\citenamefont {Manolopoulos}(1986)}]{Manolopoulos:JCP86}%
  \BibitemOpen
  \bibfield  {author} {\bibinfo {author} {\bibfnamefont {D.~E.}\ \bibnamefont
  {Manolopoulos}},\ }\bibfield  {title} {\enquote {\bibinfo {title} {An
  improved log derivative method for inelastic scattering},}\ }\href
  {https://doi.org/10.1063/1.451472} {\bibfield  {journal} {\bibinfo  {journal}
  {J. Chem. Phys.}\ }\textbf {\bibinfo {volume} {85}},\ \bibinfo {pages}
  {6425--6429} (\bibinfo {year} {1986})},\ \Eprint
  {https://arxiv.org/abs/https://doi.org/10.1063/1.451472}
  {https://doi.org/10.1063/1.451472} \BibitemShut {NoStop}%
\bibitem [{\citenamefont {Bakr}, \citenamefont {Smith},\ and\ \citenamefont
  {Patkowski}(2013)}]{3D_PES}%
  \BibitemOpen
  \bibfield  {author} {\bibinfo {author} {\bibfnamefont {B.~W.}\ \bibnamefont
  {Bakr}}, \bibinfo {author} {\bibfnamefont {D.~G.~A.}\ \bibnamefont {Smith}},\
  and\ \bibinfo {author} {\bibfnamefont {K.}~\bibnamefont {Patkowski}},\
  }\bibfield  {title} {\enquote {\bibinfo {title} {Highly accurate potential
  energy surface for the $\mathrm{He}$-$\mathrm{H}_2$ dimer},}\ }\href
  {https://doi.org/10.1063/1.4824299} {\bibfield  {journal} {\bibinfo
  {journal} {J. Chem. Phys.}\ }\textbf {\bibinfo {volume} {139}},\ \bibinfo
  {pages} {144305} (\bibinfo {year} {2013})}\BibitemShut {NoStop}%
\bibitem [{\citenamefont {Aldegunde}\ \emph {et~al.}(2005)\citenamefont
  {Aldegunde}, \citenamefont {{de Miranda}}, \citenamefont {Haigh},
  \citenamefont {Kendrick}, \citenamefont {Saez-Rabanos},\ and\ \citenamefont
  {Aoiz}}]{AMHKA:JPCA05}%
  \BibitemOpen
  \bibfield  {author} {\bibinfo {author} {\bibfnamefont {J.}~\bibnamefont
  {Aldegunde}}, \bibinfo {author} {\bibfnamefont {M.~P.}\ \bibnamefont {{de
  Miranda}}}, \bibinfo {author} {\bibfnamefont {J.~M.}\ \bibnamefont {Haigh}},
  \bibinfo {author} {\bibfnamefont {B.~K.}\ \bibnamefont {Kendrick}}, \bibinfo
  {author} {\bibfnamefont {V.}~\bibnamefont {Saez-Rabanos}},\ and\ \bibinfo
  {author} {\bibfnamefont {F.~J.}\ \bibnamefont {Aoiz}},\ }\bibfield  {title}
  {\enquote {\bibinfo {title} {How reactans polarization can be used to change
  and unravel chemical reactivity},}\ }\href
  {https://doi.org/10.1021/jp0512208} {\bibfield  {journal} {\bibinfo
  {journal} {J. Phys. Chem. A}\ }\textbf {\bibinfo {volume} {109}},\ \bibinfo
  {pages} {6200--6217} (\bibinfo {year} {2005})}\BibitemShut {NoStop}%
\bibitem [{\citenamefont {Zhou}\ and\ \citenamefont
  {Chen}(2017)}]{ICS_HD-He_Boyi}%
  \BibitemOpen
  \bibfield  {author} {\bibinfo {author} {\bibfnamefont {B.}~\bibnamefont
  {Zhou}}\ and\ \bibinfo {author} {\bibfnamefont {M.}~\bibnamefont {Chen}},\
  }\bibfield  {title} {\enquote {\bibinfo {title} {Quantum rotational
  scattering of $\mathrm{H_2}$ and its isotopologues with $\mathrm{He}$},}\
  }\href {https://doi.org/10.1080/00268976.2017.1322721} {\bibfield  {journal}
  {\bibinfo  {journal} {Mol. Phys.}\ }\textbf {\bibinfo {volume} {115}},\
  \bibinfo {pages} {2442--2450} (\bibinfo {year} {2017})}\BibitemShut {NoStop}%
\bibitem [{\citenamefont {Jambrina}\ \emph {et~al.}(2020)\citenamefont
  {Jambrina}, \citenamefont {González-Sánchez}, \citenamefont {Lara},
  \citenamefont {Menéndez},\ and\ \citenamefont {Aoiz}}]{HHF_Jambrina20}%
  \BibitemOpen
  \bibfield  {author} {\bibinfo {author} {\bibfnamefont {P.~G.}\ \bibnamefont
  {Jambrina}}, \bibinfo {author} {\bibfnamefont {L.}~\bibnamefont
  {González-Sánchez}}, \bibinfo {author} {\bibfnamefont {M.}~\bibnamefont
  {Lara}}, \bibinfo {author} {\bibfnamefont {M.}~\bibnamefont {Menéndez}},\
  and\ \bibinfo {author} {\bibfnamefont {F.~J.}\ \bibnamefont {Aoiz}},\
  }\bibfield  {title} {\enquote {\bibinfo {title} {Unveiling shape resonances
  in {H + HF} collisions at cold energies},}\ }\href
  {https://doi.org/10.1039/D0CP04885B} {\bibfield  {journal} {\bibinfo
  {journal} {Phys. Chem. Chem. Phys.}\ }\textbf {\bibinfo {volume} {22}},\
  \bibinfo {pages} {24943--24950} (\bibinfo {year} {2020})}\BibitemShut
  {NoStop}%
\bibitem [{\citenamefont {Balakrishnan}, \citenamefont {Forrey},\ and\
  \citenamefont {Dalgarno}(1997)}]{BALAKRISHNAN19971}%
  \BibitemOpen
  \bibfield  {author} {\bibinfo {author} {\bibfnamefont {N.}~\bibnamefont
  {Balakrishnan}}, \bibinfo {author} {\bibfnamefont {R.}~\bibnamefont
  {Forrey}},\ and\ \bibinfo {author} {\bibfnamefont {A.}~\bibnamefont
  {Dalgarno}},\ }\bibfield  {title} {\enquote {\bibinfo {title} {Threshold
  phenomena in ultracold atom–molecule collisions},}\ }\href
  {https://www.sciencedirect.com/science/article/pii/S0009261497010518}
  {\bibfield  {journal} {\bibinfo  {journal} {Chem. Phys. Lett.}\ }\textbf
  {\bibinfo {volume} {280}},\ \bibinfo {pages} {1--4} (\bibinfo {year}
  {1997})}\BibitemShut {NoStop}%
\bibitem [{\citenamefont {Lara}\ \emph {et~al.}(2015)\citenamefont {Lara},
  \citenamefont {Jambrina}, \citenamefont {Aoiz},\ and\ \citenamefont
  {Launay}}]{Lara-JCP15}%
  \BibitemOpen
  \bibfield  {author} {\bibinfo {author} {\bibfnamefont {M.}~\bibnamefont
  {Lara}}, \bibinfo {author} {\bibfnamefont {P.~G.}\ \bibnamefont {Jambrina}},
  \bibinfo {author} {\bibfnamefont {F.~J.}\ \bibnamefont {Aoiz}},\ and\
  \bibinfo {author} {\bibfnamefont {J.-M.}\ \bibnamefont {Launay}},\ }\bibfield
   {title} {\enquote {\bibinfo {title} {{Cold and ultracold dynamics of the
  barrierless D$^+$ + H$_2$ reaction: Quantum reactive calculations for $\sim$
  R$^{-4}$ long range interaction potential}},}\ }\href
  {https://doi.org/10.1063/1.4936144} {\bibfield  {journal} {\bibinfo
  {journal} {J. Chem. Phys.}\ }\textbf {\bibinfo {volume} {143}},\ \bibinfo
  {pages} {204305} (\bibinfo {year} {2015})}\BibitemShut {NoStop}%
\bibitem [{\citenamefont {Aldegunde}\ \emph {et~al.}(2006)\citenamefont
  {Aldegunde}, \citenamefont {Alvariño}, \citenamefont {de~Miranda},
  \citenamefont {Sáez~Rábanos},\ and\ \citenamefont
  {Aoiz}}]{Aldegunde-JCP06}%
  \BibitemOpen
  \bibfield  {author} {\bibinfo {author} {\bibfnamefont {J.}~\bibnamefont
  {Aldegunde}}, \bibinfo {author} {\bibfnamefont {J.~M.}\ \bibnamefont
  {Alvariño}}, \bibinfo {author} {\bibfnamefont {M.~P.}\ \bibnamefont
  {de~Miranda}}, \bibinfo {author} {\bibfnamefont {V.}~\bibnamefont
  {Sáez~Rábanos}},\ and\ \bibinfo {author} {\bibfnamefont {F.~J.}\
  \bibnamefont {Aoiz}},\ }\bibfield  {title} {\enquote {\bibinfo {title}
  {{Mechanism and control of the F+H2 reaction at low and ultralow collision
  energies}},}\ }\href {https://doi.org/10.1063/1.2212418} {\bibfield
  {journal} {\bibinfo  {journal} {J. Chem. Phys.}\ }\textbf {\bibinfo {volume}
  {125}},\ \bibinfo {pages} {133104} (\bibinfo {year} {2006})}\BibitemShut
  {NoStop}%
\bibitem [{\citenamefont {Jambrina}\ \emph {et~al.}(2015)\citenamefont
  {Jambrina}, \citenamefont {Herr\'aez-Aguilar}, \citenamefont {Aoiz},
  \citenamefont {Sneha}, \citenamefont {Jankunas},\ and\ \citenamefont
  {Zare}}]{JHASJZ:NC15}%
  \BibitemOpen
  \bibfield  {author} {\bibinfo {author} {\bibfnamefont {P.~G.}\ \bibnamefont
  {Jambrina}}, \bibinfo {author} {\bibfnamefont {D.}~\bibnamefont
  {Herr\'aez-Aguilar}}, \bibinfo {author} {\bibfnamefont {F.~J.}\ \bibnamefont
  {Aoiz}}, \bibinfo {author} {\bibfnamefont {M.}~\bibnamefont {Sneha}},
  \bibinfo {author} {\bibfnamefont {J.}~\bibnamefont {Jankunas}},\ and\
  \bibinfo {author} {\bibfnamefont {R.~N.}\ \bibnamefont {Zare}},\ }\bibfield
  {title} {\enquote {\bibinfo {title} {Quantum interference between {H + D$_2$}
  quasiclassical reaction mechanisms},}\ }\href
  {https://doi.org/10.1038/nchem.2295} {\bibfield  {journal} {\bibinfo
  {journal} {Nat. Chem.}\ }\textbf {\bibinfo {volume} {7}},\ \bibinfo {pages}
  {661} (\bibinfo {year} {2015})}\BibitemShut {NoStop}%
\bibitem [{\citenamefont {Jambrina}, \citenamefont {Men\'endez},\ and\
  \citenamefont {Aoiz}(2018)}]{JMA:CS18}%
  \BibitemOpen
  \bibfield  {author} {\bibinfo {author} {\bibfnamefont {P.~G.}\ \bibnamefont
  {Jambrina}}, \bibinfo {author} {\bibfnamefont {M.}~\bibnamefont
  {Men\'endez}},\ and\ \bibinfo {author} {\bibfnamefont {F.~J.}\ \bibnamefont
  {Aoiz}},\ }\bibfield  {title} {\enquote {\bibinfo {title} {Angular momentum
  scattering angle quantum correlation: a generalized deflection function},}\
  }\href {https://doi.org/10.1039/C7SC05489K} {\bibfield  {journal} {\bibinfo
  {journal} {Chem. Sci.}\ }\textbf {\bibinfo {volume} {9}},\ \bibinfo {pages}
  {4837} (\bibinfo {year} {2018})}\BibitemShut {NoStop}%
\bibitem [{\citenamefont {Amarasinghe}\ \emph {et~al.}()\citenamefont
  {Amarasinghe}, \citenamefont {Perera}, \citenamefont {Li}, \citenamefont
  {Zuo}, \citenamefont {Besemer}, \citenamefont {van~der Avoird}, \citenamefont
  {Groenenboom}, \citenamefont {Guo},\ and\ \citenamefont
  {Suits}}]{Suits_NS22}%
  \BibitemOpen
  \bibfield  {author} {\bibinfo {author} {\bibfnamefont {C.}~\bibnamefont
  {Amarasinghe}}, \bibinfo {author} {\bibfnamefont {C.~A.}\ \bibnamefont
  {Perera}}, \bibinfo {author} {\bibfnamefont {H.}~\bibnamefont {Li}}, \bibinfo
  {author} {\bibfnamefont {J.}~\bibnamefont {Zuo}}, \bibinfo {author}
  {\bibfnamefont {M.}~\bibnamefont {Besemer}}, \bibinfo {author} {\bibfnamefont
  {A.}~\bibnamefont {van~der Avoird}}, \bibinfo {author} {\bibfnamefont
  {G.~C.}\ \bibnamefont {Groenenboom}}, \bibinfo {author} {\bibfnamefont
  {H.}~\bibnamefont {Guo}},\ and\ \bibinfo {author} {\bibfnamefont {A.~G.}\
  \bibnamefont {Suits}},\ }\bibfield  {title} {\enquote {\bibinfo {title}
  {Collision-induced spin-orbit relaxation of highly vibrationally excited no
  near 1 k},}\ }\href {https://doi.org/https://doi.org/10.1002/ntls.20210074}
  {\bibfield  {journal} {\bibinfo  {journal} {Nat. Sci.}\ }\textbf {\bibinfo
  {volume} {2}},\ \bibinfo {pages} {e20210074}}\BibitemShut {NoStop}%
\bibitem [{\citenamefont {Garberoglio}, \citenamefont {Patkowski},\ and\
  \citenamefont {Harvey}(2014)}]{Garberoglio_HeH2}%
  \BibitemOpen
  \bibfield  {author} {\bibinfo {author} {\bibfnamefont {G.}~\bibnamefont
  {Garberoglio}}, \bibinfo {author} {\bibfnamefont {K.}~\bibnamefont
  {Patkowski}},\ and\ \bibinfo {author} {\bibfnamefont {A.~H.}\ \bibnamefont
  {Harvey}},\ }\bibfield  {title} {\enquote {\bibinfo {title} {Fully quantum
  cross second virial coefficients for the three-dimensional {H}e--{H}$_2$
  pair},}\ }\href {https://doi.org/10.1007/s10765-014-1729-7} {\bibfield
  {journal} {\bibinfo  {journal} {Int. J. Thermophys}\ }\textbf {\bibinfo
  {volume} {35}},\ \bibinfo {pages} {1435--1449} (\bibinfo {year}
  {2014})}\BibitemShut {NoStop}%
\end{thebibliography}%


\begin{thebibliography}{6}%
\makeatletter
\providecommand \@ifxundefined [1]{%
 \@ifx{#1\undefined}
}%
\providecommand \@ifnum [1]{%
 \ifnum #1\expandafter \@firstoftwo
 \else \expandafter \@secondoftwo
 \fi
}%
\providecommand \@ifx [1]{%
 \ifx #1\expandafter \@firstoftwo
 \else \expandafter \@secondoftwo
 \fi
}%
\providecommand \natexlab [1]{#1}%
\providecommand \enquote  [1]{``#1''}%
\providecommand \bibnamefont  [1]{#1}%
\providecommand \bibfnamefont [1]{#1}%
\providecommand \citenamefont [1]{#1}%
\providecommand \href@noop [0]{\@secondoftwo}%
\providecommand \href [0]{\begingroup \@sanitize@url \@href}%
\providecommand \@href[1]{\@@startlink{#1}\@@href}%
\providecommand \@@href[1]{\endgroup#1\@@endlink}%
\providecommand \@sanitize@url [0]{\catcode `\\12\catcode `\$12\catcode
  `\&12\catcode `\#12\catcode `\^12\catcode `\_12\catcode `\%12\relax}%
\providecommand \@@startlink[1]{}%
\providecommand \@@endlink[0]{}%
\providecommand \url  [0]{\begingroup\@sanitize@url \@url }%
\providecommand \@url [1]{\endgroup\@href {#1}{\urlprefix }}%
\providecommand \urlprefix  [0]{URL }%
\providecommand \Eprint [0]{\href }%
\providecommand \doibase [0]{https://doi.org/}%
\providecommand \selectlanguage [0]{\@gobble}%
\providecommand \bibinfo  [0]{\@secondoftwo}%
\providecommand \bibfield  [0]{\@secondoftwo}%
\providecommand \translation [1]{[#1]}%
\providecommand \BibitemOpen [0]{}%
\providecommand \bibitemStop [0]{}%
\providecommand \bibitemNoStop [0]{.\EOS\space}%
\providecommand \EOS [0]{\spacefactor3000\relax}%
\providecommand \BibitemShut  [1]{\csname bibitem#1\endcsname}%
\let\auto@bib@innerbib\@empty
\bibitem [{\citenamefont {Aldegunde}\ \emph {et~al.}(2005)\citenamefont
  {Aldegunde}, \citenamefont {{de Miranda}}, \citenamefont {Haigh},
  \citenamefont {Kendrick}, \citenamefont {Saez-Rabanos},\ and\ \citenamefont
  {Aoiz}}]{AMHKA:JPCA05}%
  \BibitemOpen
  \bibfield  {author} {\bibinfo {author} {\bibfnamefont {J.}~\bibnamefont
  {Aldegunde}}, \bibinfo {author} {\bibfnamefont {M.~P.}\ \bibnamefont {{de
  Miranda}}}, \bibinfo {author} {\bibfnamefont {J.~M.}\ \bibnamefont {Haigh}},
  \bibinfo {author} {\bibfnamefont {B.~K.}\ \bibnamefont {Kendrick}}, \bibinfo
  {author} {\bibfnamefont {V.}~\bibnamefont {Saez-Rabanos}},\ and\ \bibinfo
  {author} {\bibfnamefont {F.~J.}\ \bibnamefont {Aoiz}},\ }\bibfield  {title}
  {\enquote {\bibinfo {title} {How reactans polarization can be used to change
  and unravel chemical reactivity},}\ }\href
  {https://doi.org/10.1021/jp0512208} {\bibfield  {journal} {\bibinfo
  {journal} {J. Phys. Chem. A}\ }\textbf {\bibinfo {volume} {109}},\ \bibinfo
  {pages} {6200--6217} (\bibinfo {year} {2005})}\BibitemShut {NoStop}%
\bibitem [{\citenamefont {Blum}(2012)}]{Blum2012}%
  \BibitemOpen
  \bibfield  {author} {\bibinfo {author} {\bibfnamefont {K.}~\bibnamefont
  {Blum}},\ }\href@noop {} {\emph {\bibinfo {title} {Density matrix theory and
  applications}}},\ \bibinfo {edition} {3rd}\ ed.,\ Vol.~\bibinfo {volume}
  {64}\ (\bibinfo  {publisher} {Springer Science \& Business Media, Heidelberg,
  Dordrecht, London, New York},\ \bibinfo {year} {2012})\BibitemShut {NoStop}%
\bibitem [{\citenamefont {Zhou}\ \emph
  {et~al.}(2021{\natexlab{a}})\citenamefont {Zhou}, \citenamefont {Perreault},
  \citenamefont {Mukherjee},\ and\ \citenamefont {Zare}}]{sarp_hed2}%
  \BibitemOpen
  \bibfield  {author} {\bibinfo {author} {\bibfnamefont {H.}~\bibnamefont
  {Zhou}}, \bibinfo {author} {\bibfnamefont {W.~E.}\ \bibnamefont {Perreault}},
  \bibinfo {author} {\bibfnamefont {N.}~\bibnamefont {Mukherjee}},\ and\
  \bibinfo {author} {\bibfnamefont {R.~N.}\ \bibnamefont {Zare}},\ }\bibfield
  {title} {\enquote {\bibinfo {title} {Shape resonance determined from angular
  distribution in {D}$_2$ (v=2, j=2) + {H}e $\to$ {D}$_2$ (v=2, j=0) + {H}e
  cold scattering},}\ }\href {https://doi.org/10.1063/5.0045087} {\bibfield
  {journal} {\bibinfo  {journal} {J. Chem. Phys}\ }\textbf {\bibinfo {volume}
  {154}},\ \bibinfo {pages} {104309} (\bibinfo {year}
  {2021}{\natexlab{a}})}\BibitemShut {NoStop}%
\bibitem [{\citenamefont {Zhou}\ \emph
  {et~al.}(2021{\natexlab{b}})\citenamefont {Zhou}, \citenamefont {Perreault},
  \citenamefont {Mukherjee},\ and\ \citenamefont {Zare}}]{sarp_hed2_science}%
  \BibitemOpen
  \bibfield  {author} {\bibinfo {author} {\bibfnamefont {H.}~\bibnamefont
  {Zhou}}, \bibinfo {author} {\bibfnamefont {W.~E.}\ \bibnamefont {Perreault}},
  \bibinfo {author} {\bibfnamefont {N.}~\bibnamefont {Mukherjee}},\ and\
  \bibinfo {author} {\bibfnamefont {R.~N.}\ \bibnamefont {Zare}},\ }\bibfield
  {title} {\enquote {\bibinfo {title} {Quantum mechanical double slit for
  molecular scattering},}\ }\href
  {https://www.science.org/doi/10.1126/science.abl4143} {\bibfield  {journal}
  {\bibinfo  {journal} {Science}\ }\textbf {\bibinfo {volume} {374}},\ \bibinfo
  {pages} {960--964} (\bibinfo {year} {2021}{\natexlab{b}})}\BibitemShut
  {NoStop}%
\bibitem [{\citenamefont {W.~E.~Perreault}(2018)}]{ZareCP18}%
  \BibitemOpen
  \bibfield  {author} {\bibinfo {author} {\bibfnamefont {R.~N.~Z.}\
  \bibnamefont {W.~E.~Perreault}, \bibfnamefont {N.~Mukherjee}},\ }\bibfield
  {title} {\enquote {\bibinfo {title} {Supersonic beams of mixed gases: A
  method for studying cold collisions},}\ }\href
  {https://www.sciencedirect.com/science/article/pii/S0301010417309977?via%3Dihub}
  {\bibfield  {journal} {\bibinfo  {journal} {Chem. Phys.}\ }\textbf {\bibinfo
  {volume} {514}},\ \bibinfo {pages} {150--153} (\bibinfo {year}
  {2018})}\BibitemShut {NoStop}%
\bibitem [{\citenamefont {Thibault}\ \emph {et~al.}(2017)\citenamefont
  {Thibault}, \citenamefont {Patkowski}, \citenamefont {\.{Z}uchowski},
  \citenamefont {J\'{o}\'{z}wiak}, \citenamefont {Ciury\l{}o},\ and\
  \citenamefont {Wcis\l{}o}}]{BSP3}%
  \BibitemOpen
  \bibfield  {author} {\bibinfo {author} {\bibfnamefont {F.}~\bibnamefont
  {Thibault}}, \bibinfo {author} {\bibfnamefont {K.}~\bibnamefont {Patkowski}},
  \bibinfo {author} {\bibfnamefont {P.~S.}\ \bibnamefont {\.{Z}uchowski}},
  \bibinfo {author} {\bibfnamefont {H.}~\bibnamefont {J\'{o}\'{z}wiak}},
  \bibinfo {author} {\bibfnamefont {R.}~\bibnamefont {Ciury\l{}o}},\ and\
  \bibinfo {author} {\bibfnamefont {P.}~\bibnamefont {Wcis\l{}o}},\ }\bibfield
  {title} {\enquote {\bibinfo {title} {Rovibrational line-shape parameters for
  $\mathrm{H}_2$ in $\mathrm{He}$ and new $\mathrm{H}_2$-$\mathrm{He}$
  potential energy surface},}\ }\href
  {https://doi.org/https://doi.org/10.1016/j.jqsrt.2017.08.014} {\bibfield
  {journal} {\bibinfo  {journal} {J. Quant. Spectrosc. Radiat. Transf.}\
  }\textbf {\bibinfo {volume} {202}},\ \bibinfo {pages} {308} (\bibinfo {year}
  {2017})}\BibitemShut {NoStop}%
\end{thebibliography}%
\end{document}



\title{Supplementary Information for: The role of low energy resonances in the stereodynamics of cold He+D$_2$ collisions}

\author{Pablo G. Jambrina}
\affiliation{Departamento de Qu\'imica F\'isica, University of Salamanca, Salamanca
37008, Spain. e-mail: pjambrina@usal.es}
\author{Masato Morita}
\affiliation{
Department of Chemistry and Biochemistry, University of Nevada, Las Vegas, Nevada 89154,
USA
}%
\author{James F. E. Croft}
\affiliation{Department of Physics, University of Otago, Dunedin
9054, New Zealand. e-mail: j.croft@otago.ac.nz}
\affiliation{Dodd-Walls Centre for Photonic and Quantum
Technologies, Dunedin 9054, New Zealand.}
\author{F. Javier Aoiz}
\affiliation{Departamento de Qu\'imica F\'isica, Universidad Complutense, Madrid
28040, Spain. e-mail:aoiz@quim.ucm.es}
\author{Naduvalath Balakrishnan}%
\email{naduvala@unlv.nevada.edu}
\affiliation{
Department of Chemistry and Biochemistry, University of Nevada, Las Vegas, Nevada 89154,
USA
}%

\date{\today}

\maketitle

\tableofcontents

\section{Deduction of X-SARP polarization parameters}

Following Ref.~\citenum{AMHKA:JPCA05}, the probability density function (PDF) that describes the spatial distribution of the internuclear axis following SARP excitation is given by:
\begin{equation}\label{eq:extrinsic}
P(\theta_r,\phi_r) = \sum_{k=0}^{2j} \sum_{q=-k}^{q=k} \frac{2 k +1}{4 \pi} a^{(k)}_q
\langle j 0, k 0 | j 0 \rangle C^*_{k,q} (\theta_r,\phi_r),
\end{equation}
where $\theta_r$, and $\phi_r$ are the polar and azimuthal angles that specify the direction of the
D$_2$ internuclear axis with respect to the scattering frame, $C_{k,q}$ is the modified spherical
harmonic, $\langle .. | .. \rangle$ is a Clebsch-Gordan coefficient, and $a^{(k)}_q$ are the
extrinsic (preparation) polarization parameters in the $\bm k-\bm k'$ frame.

The $a^{(k)}_q$ can be obtained from the extrinsic polarization parameters in the laboratory-fixed frame, ${A}^{(k)}_Q$, following the general expression,
\begin{equation}\label{eq:extrinsicpp2}
a^{(k)}_q  = \sum_Q [D^{k}_{q Q}(\alpha, \beta, \gamma)]^* {A}^{(k)}_Q \,.
\end{equation}

On the other hand, the polarization parameters (PP) $A^{(k)}_Q$ can be derived from the density matrix as:\cite{Blum2012}
\begin{align}\label{Akqrho}
A^{(k)}_Q = \sum_{m_1} \, \sum_{m_2} \, \langle j m_1 |\rho| j m_2 \rangle  \,\, \langle  j m_1 \, k q |j m_2 \rangle
\end{align}
In the case of a pure state resulting from the coherent superposition of $m$=1 and $m=-1$ states, such as that prepared via X-SARP:
\begin{equation}
|\psi \rangle= \frac{1}{\sqrt{2}}\, [|j \, m=-1\rangle - | j\, m=+1\rangle]
\end{equation}
and the density operator (omitting $j$ for brevity)
\begin{equation}\label{density}
\rho= |\psi \rangle\langle \psi| = \frac{1}{2}\, \Big[ | +1 \rangle\langle +1 |  +
| -1 \rangle\langle -1 |  - |+1\rangle \langle -1| - |-1\rangle \langle +1| \Big]
\end{equation}
Hence the density matrix elements are $\rho_{ij}= \langle m_i | \rho | m_j \rangle =+\frac{1}{2}$
if $m_i=m_j$ and $-\frac{1}{2}$ if  $m_i \ne  m_j$ with $m_i, ~m_j= \pm 1$.

Plugging the density matrix from Eq. \eqref{density} into Eq.~\eqref{Akqrho}, leads to the resulting polarization parameters:
\begin{align}\label{PP}
&A^{(0)}_0=1 \\
&A^{(2)}_0 = - \sqrt{\frac{1}{14}}; \qquad A^{(2)}_2 =A^{(2)}_{-2}= - \frac{1}{2} \, \sqrt{\frac{3}{7}} \\
&A^{(4)}_0 = - \frac{2}{3}\, \sqrt{\frac{2}{7}};\qquad A^{(4)}_2 =A^{(4)}_{-2}= + \frac{1}{3} \, \sqrt{\frac{5}{7}} \label{PPP}
\end{align}

The laboratory-fixed frame for a X-SARP preparation is chosen so that the $\rm Z$ axis is parallel
to the pump laser polarization and $\rm X$ the axis lies along the Stokes laser polarization.  The
distribution of internuclear axis in this frame can be written as:
\begin{align}\label{dist1}
P(\Theta_r, \Phi_r) =\sum_k^{2j}  \sum_{q=-k}^{+k}  \, \left(\frac{2k+1}{4\pi}\right)\, \langle j
0 \,\, k 0 | j 0 \rangle \, A^{(k)}_q C_{k q}^*(\Theta_r, \Phi_r)
\end{align}
where $\Theta_r$ and $\Phi_r$  define the direction of the internuclear axis in the Pump-Stokes
frame. In the case that laboratory frame and scattering frame would coincide, this equation would result of setting  $\alpha=\beta=\gamma=0$ in Eq.~\eqref{eq:extrinsicpp2}, and introducing the resulting $a^{(k)}_q$ in Eq.~\eqref{eq:extrinsic}.

In the particular case of $j$=2, Eq.~\eqref{dist1} transforms to
\begin{align}\label{dist2}
P(\Theta_r, \Phi_r)&=\frac{1}{4\pi} \, \Big\{1 + 5 \times \,\, \langle 2 0 \,\,2 0 | 2 0 \rangle  \Big[ A^{(2)}_0 \, C_{2 0}(\Theta_r, 0) +
 2 \, A^{(2)}_2 C_{2 2}(\Theta_r, 0) \cos2\Phi_r \Big] + \nonumber \\
&9 \times \,\,\langle 2 0 \,4 0 | 2 0 \rangle  \Big[ A^{(4)}_0 \, C_{4 0}(\Theta_r, 0) +
 2 \, A^{(4)}_2 C_{4 2}(\Theta_r, 0) \cos2\Phi_r \Big] \Big\},
\end{align}
where we have used that  $\langle j 0 \,\,k 0 | j 0 \rangle=0$ for odd values of $k$.

Inserting the PPs from Eq.~\eqref{PP}-\eqref{PPP} in Eq.~\eqref{dist2}, the (polar) internuclear axis distribution can be written
as
\begin{equation}
P(\Theta_r, \Phi_r)= \frac{15}{4\pi} \, \sin^2\Theta_r \cos^2\Theta_r \cos^2\Phi_r\, ,
\end{equation}
that is the square of the spherical harmonic  $d_{\rm XZ}$.

\section{Deduction of X-SARP DCS}

For an experiment in which the pump and Stokes laser are parallel to each other (such as in V-SARP,
H-SARP and 45-SARP), we could choose a reference frame whose $\rm Z$ axis lies along the pump
and Stokes laser, and whose $\rm X$ and $\rm Y$ axis are arbitrarily chosen. In a SARP experiment,
the wave function that represents the prepared state with respect to this frame is $| j \, m
\rangle=|2 \, 0\rangle$, such that its distribution of internuclear axis is aligned along $\rm Z$.

However, we are interested in the scattering events referred to the scattering frame, in
which the $z$ axis is along the initial relative velocity, $\bm k$, and the $xz$ plane is defined
by $\bm k$ and the final relative velocity $\bm k'$. With respect to the scattering frame, the direction of
the pump and Stokes polarization vectors is defined by the  polar angle  $\beta$ and the azimuthal
angle $\alpha$. To describe the wave function in the scattering frame, it is necessary  to rotate it in the $\bm k-\bm k'$
frame according to:
\begin{align}
|\psi\rangle &= \hat{R}(\phi, \beta, \chi=0) |2\, 0\rangle= \sum_m D^j_{m0}(\alpha, \beta, \gamma=0) |2 \, m \rangle=
\sum_m C_{jm}(\beta, 0) e^{-im\alpha} |2\, m \rangle= \nonumber \\
&C_{20}(\beta, 0) |2 \,0\rangle + C_{2\,1}(\beta, 0) \Big[e^{-i\alpha} |2\, 1\rangle - e^{i\alpha} |2\, -1\rangle   \Big] +
C_{2\,2}(\beta, 0) \Big[ e^{-2i\alpha} |2 \, 2\rangle + e^{2i\alpha} |2\, -2\rangle \Big]
\end{align}

Let us compare $|\psi\rangle$ for $\beta$ and $( \pi - \beta)$ angles:
\begin{equation}
|\psi_{\beta}\rangle= C_{20}(\beta, 0) |2 \, 0\rangle + C_{2\,1}(\beta, 0) \Big[ e^{-i\alpha} |2 \, 1\rangle - e^{i\alpha} |2\, -1\rangle \Big] +
C_{2\,2}(\beta, 0) \Big[ e^{-2i\alpha} |2 \, 2\rangle + e^{2i\alpha} |2\, -2\rangle \Big] \label{beta1}
\end{equation}
and since $C_{jm}(\pi-\beta, 0)=(-1)^{j+m}\,C_{jm}(\beta,0)$,
\begin{align}
|\psi_{\pi -\beta}\rangle = C_{20}(\beta, 0) |2 \,0\rangle - & C_{2\,1}(\beta, 0)
\Big[ e^{-i\alpha} |2 \, 1\rangle - e^{i\alpha} |2\, -1\rangle \Big] + \nonumber \\
& C_{2\,2}(\beta, 0) \Big[ e^{-2i\alpha} |2 \, 2\rangle + e^{2i\alpha} |2\, -2\rangle \Big] \label{beta2}
\end{align}
It is then possible to define a normalized function $|{\tilde \psi} \rangle$ superposition of
$|\psi_{\beta}\rangle$ and $|\psi_{\pi-\beta}\rangle$ (0 < $\beta$ < $\pi/2$)
\begin{equation}
|{\tilde\psi_X} \rangle= \frac{1}{\sqrt{2}} (|\psi_{\beta}\rangle - |\psi_{\pi -\beta}\rangle ) =
\sqrt{2}\,\, C_{21}(\beta, 0) \, \Big[( e^{-i\alpha}\, |2\, 1 \rangle  - e^{i\alpha} \, |2\, -1\rangle \Big]
\label{psihat0}
\end{equation}
that only includes contributions from $m$=$\pm$1. For the particular case of $\beta= \pi/4$, and writing $|\psi_+ \rangle=
|\psi_{\beta}\rangle$ and $|\psi_-\rangle=|\psi_{\pi -\beta}\rangle$,
\begin{equation}
|{\tilde\psi_X} \rangle= \frac{1}{\sqrt{2}} (|\psi_+\rangle - |\psi_-\rangle ) =
\frac{\sqrt{3}}{2} \,\Big( e^{i\alpha}\, |2\,-1\rangle -  e^{-i\alpha}\, |2 \, 1 \rangle \Big)
\label{psihat}
\end{equation}

Now let us consider a preparation with cross-polarized pump (along $\rm Z$=$z$=$\bm { k}$) and
Stokes (along an arbitrary $\rm X$ direction perpendicular to $z$) laser pulses.  Notice that the
$\rm Z$ axis is made to coincide with the time-of-flight axis in the experiments by Zhou {\em
et al}.\cite{sarp_hed2,sarp_hed2_science} In this case, it is possible to obtain a pure state as a superposition of $|2 -1\rangle$
and $|2 1\rangle$:
\begin{equation}
|\psi_X \rangle= \frac{1}{\sqrt{2}} \Big(|2 -1\rangle  - |2 \,1\rangle \Big)
 =d_{\rm XZ}= \frac{1}{\sqrt{2}} \big[-Y_{2\, 1}(\theta_r,\phi_r) + Y_{2-1}(\theta_r,\phi_r) \big] \label{psi}\,,
\end{equation}
As shown in the main text, when referring this state to the $\bm k-\bm k'$ frame:
\begin{equation}
|\psi_X \rangle= \frac{1}{\sqrt{2}} (e^{i\alpha} |2 -1\rangle  - e^{-i\alpha} |2 \,1\rangle)
\label{psi}
\end{equation}
Note that $\alpha$ is a phase factor, and the effect of changing $\alpha$ is the rotation around
Z.

Comparing Eq.~\eqref{psi} with Eq.~\eqref{psihat}:
\begin{equation}
|\psi_X \rangle= \sqrt{\frac{2}{3}}\, |{\tilde\psi_X} \rangle =\frac{1}{\sqrt{3}}\big( |\psi_+ \rangle - |\psi_-\rangle \big) \label{comp}
\end{equation}
Let us use the expression of $|\psi_X \rangle$ in Eq.~\eqref{psi} to determine
the scattering amplitude:
\begin{equation}\label{fx}
F_X(\theta, \phi) = \frac{1}{\sqrt{2}} \big(F_{0-1} e^{-i\phi}  - F_{01} e^{i\phi}\big)
\end{equation}
where the scattering amplitude $F_{m' m}\equiv f_{j'm'\, jm}$, and $\phi= \phi_r-\alpha$. Since
$F_{0-1}= - F_{01}$, Eq.~\eqref{fx} can be written as
\begin{equation}
|F_X(\theta, \phi)|^2 = \frac{1}{2} |F_{0-1} e^{-i\phi} - F_{01} e^{i\phi}|^2= 2 |F_{01} \cos \phi|^2,
\end{equation}
Since the experiments has been carried out under conditions of azimuthal symmetry, by integrating
over $\phi$:
\begin{align} \label{IX}
I_X(\theta)= \frac{d \sigma}{d\theta}=\sin\theta \int_0^{2\pi} \,\, \frac{d \sigma}{d\omega} d\phi   =
\sin\theta \int_0^{2\pi} |F_X(\theta, \phi)|^2 d\phi= 2\pi \, \sin \theta\, |F_{01}|^2
\end{align}
which is Eq. 15 of the main text.

Going back to Eq.~\eqref{psihat}, it is possible to calculate the scattering amplitudes associated to $|\psi_+ \rangle$ and $|\psi_- \rangle$,
\begin{align}
F_+(\theta, \phi) &=\frac{1}{4} F_{00}(\theta) + \sqrt{\frac{3}{2}} \, F_{0 \,-1}(\theta)\,\cos \phi  +
\sqrt{\frac{3}{8}} F_{0 \, 2}(\theta) \cos2\phi  \label{f+} \\
F_-(\theta, \phi) &=\frac{1}{4} F_{00}(\theta) - \sqrt{\frac{3}{2}} \, F_{0 \,-1}(\theta)\,\cos \phi  +
\sqrt{\frac{3}{8}} F_{0 \, 2}(\theta) \cos2\phi  \label{f-}
\end{align}
and
\begin{equation}
F_X(\theta, \phi) = \frac{1}{\sqrt{3}} \, \big[ F_{+}(\theta) - F_{-}(\theta) \big]
\end{equation}
Squaring the scattering amplitude $F_X(\theta, \phi)$,
%
%
\begin{equation}
|F_X(\theta, \phi)|^2 = \frac{1}{3} \Big[ |F_{+}(\theta)|^2 +|F_-(\theta)|^2 \underbrace{-
F_{+}(\theta)F_-^*(\theta) - F_{+}^*(\theta) F_-(\theta)}_{|F_{\rm int}(\theta, \phi)|^2} \Big]
\end{equation}
and the DCSs resulting from the respective expressions of azimuthally integrated
\begin{equation}
I_+(\theta)=  \sin\theta \int_0^{2\pi} |F_+(\theta, \phi)|^2 d\phi=
\sin\theta  \Big[ \frac{\pi}{8} |F_{00}(\theta)|^2  + \frac{3\pi}{2}\,|F_{01}(\theta)|^2 +
\frac{3\pi}{8}|F_{02}(\theta)|^2 \Big]= I_-(\theta)
\end{equation}
\begin{align}
I_{\rm int}(\theta)=   \sin\theta \int_0^{2\pi}  \Big[-F_{+}(\theta)F_-^*(\theta) - F_{+}^*(\theta) F_-(\theta) \Big] d\phi=
- \sin\theta \Big[ \frac{\pi}{4} |F_{00}|^2  {\bf -}  3\pi |F_{01}|^2 +
\frac{3\pi}{4}|F_{02}|^2 \Big]
\end{align}
which are Eq. 14 and 18 of the main text, so that
\begin{equation}\label{xsarpintsi}
I_{X}(\theta) = \frac{2}{3} I_{+}(\theta) + \frac{1}{3} I_{\rm int}(\theta)\,,
\end{equation}
(Eq. 17 of the main text).

\section{Determination of the Collision Energy distributions}

To calculate the velocity-averaged differential rate coefficients shown in Fig. 4 and 5 of the main manuscript, it is necessary to weight $I_{\beta}(\theta)$ by the experimental collision energy distribution. This energy distribution was computed as follows:
\begin{itemize}
  \item Velocities of He and D$_2$ are  randomly sampled according to the experimental Gaussian laboratory velocity distributions whose peak velocities were  2067 m/s for He and 2061 m/s for D$_2$, corresponding to the respective relative velocities of 0 m/s  and -6 m/s,  shown  in the SI of Ref.~\citenum{sarp_hed2}.\cite{ZareCP18}

  \item The relative velocity, $v_{\rm rel}$ is calculated as:
  \begin{equation}
    v_{\rm rel}^2 = v_{\rm He}^2 + v_{\rm D2}^2 - 2 v_{\rm He} v_{\rm D2} \cos \chi
  \end{equation}
 where $\chi$ is the  crossing angle between $\bm v_{\rm He}$ and $\bm v_{\rm D2}$, and it is sampled according to the beam divergence (experimentally 12 mRad) and the azimuthal symmetry of the experiment.  If the divergence is neglected ({\em i.e.}, if we were considering a 1D collision energy distribution),  only the component $v_z$ components of $\bm v_{He}$ and $\bm v_{D2}$ are used for the calculation of the collision energy distribution. In a 1D distribution, $\cos \chi = 1$, so $v_{\rm rel} = v_{\rm He} - v_{\rm D2}$.

 \item $E_{\rm coll}$ is calculated from $v_{\rm rel}$. to ensure the  convergence in the collision energy distribution, the sampling is repeated $5 \times 10^7$ times.
\end{itemize}


\clearpage

\section{Supplementary Figures}

\begin{figure}[h!]
\begin{center}
\includegraphics[width=1.0\linewidth]{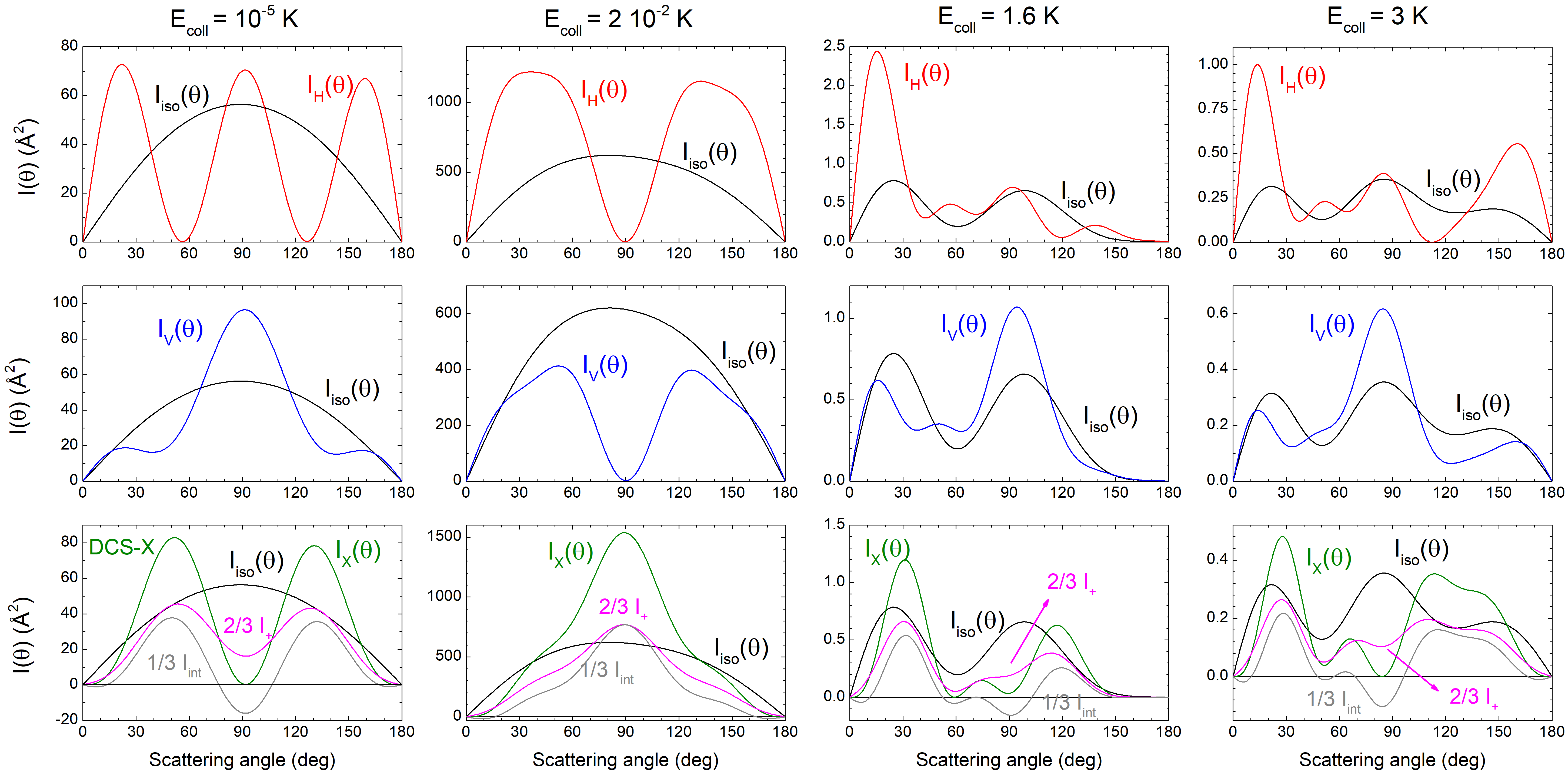}
\end{center}
\caption{Differential Cross Sections for the D$_2$ ($v=2,j=2 \to v'=2,j'=0$) by collisions with
He for different initial preparations of the D$_2$ rotational state and four different energies:
10$^{-5}$, $2 \cdot 10^{-2}$, 1.6, and 3 K. In the top panels, the results for the H-SARP
preparations are shown, while the middle panels show the results for V-SARP. In the bottom panels,
the results for the X-SARP are shown along with the results for 45-SARP (I$_+$), and
DCS$_{\rm int}$. In all panels the isotropic DCS is shown for the sake of comparison.
The DCSs are multiplied by $\sin\theta$ to facilitate comparison with experimental differential flux.
}
\label{fig:DCSenergysin}
\end{figure}

\begin{figure}[b!]
\begin{center}
\includegraphics[width=1.0\linewidth]{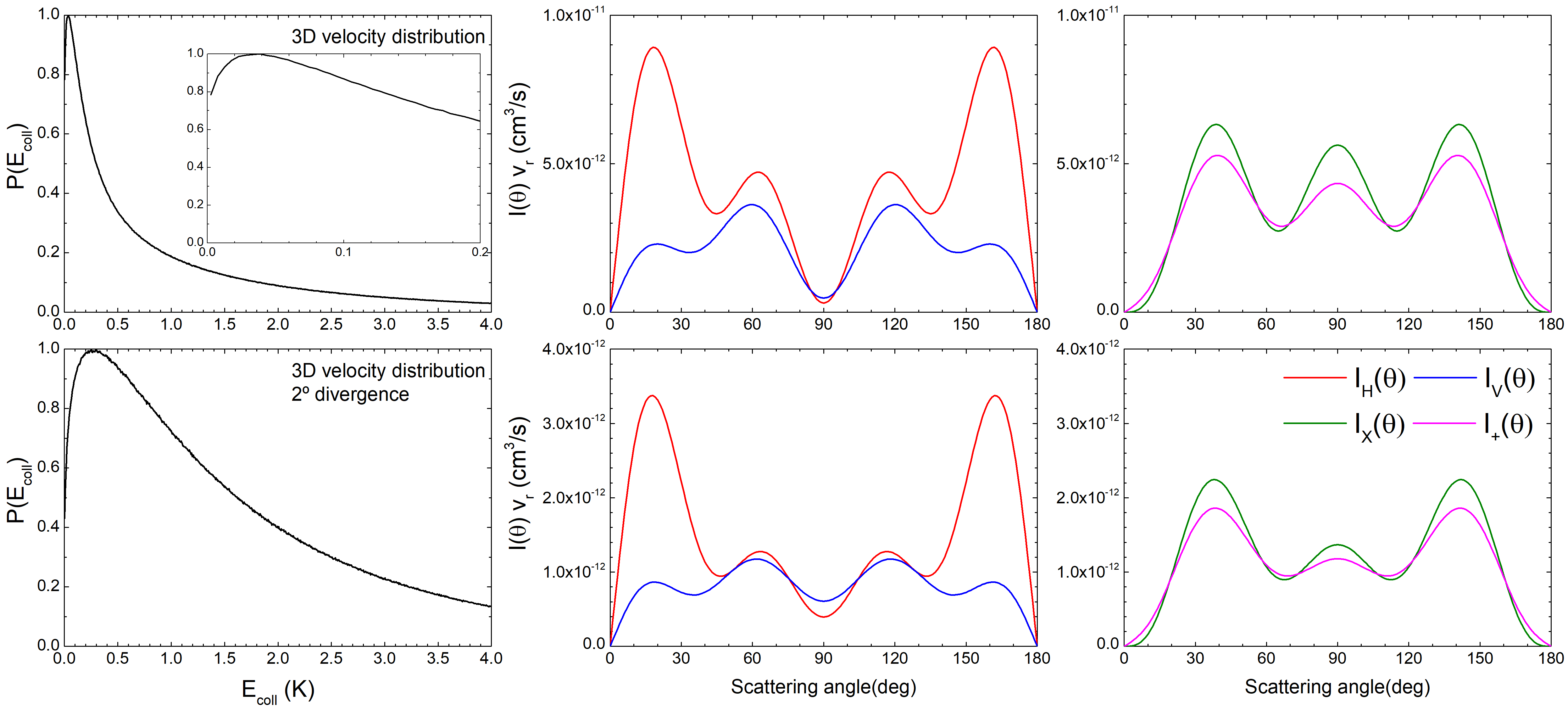}
\end{center}
\caption{Velocity-averaged differential rate coefficients for D$_2$ ($v=2,j=2 \to v'=2,j'=0$) by
collisions with He for the H-SARP,  V-SARP, X-SARP, and 45-SARP preparations of the initial
D$_2$ orientations as functions of the scattering angle. Results are shown for two different
collision energy distributions, one assuming a 3D distribution with a 12mRad=0.7$^{\circ}$ divergence as in the experiment (top panel), and another
assuming a 3D distribution where the beam has a larger divergence (bottom panel). Both P(E$_{\rm coll}$) are normalized to 1 at the maximum.
}
\label{fig:DCS_aver2}
\end{figure}

\begin{figure}[h!]
\begin{center}
\includegraphics[width=1.0\linewidth]{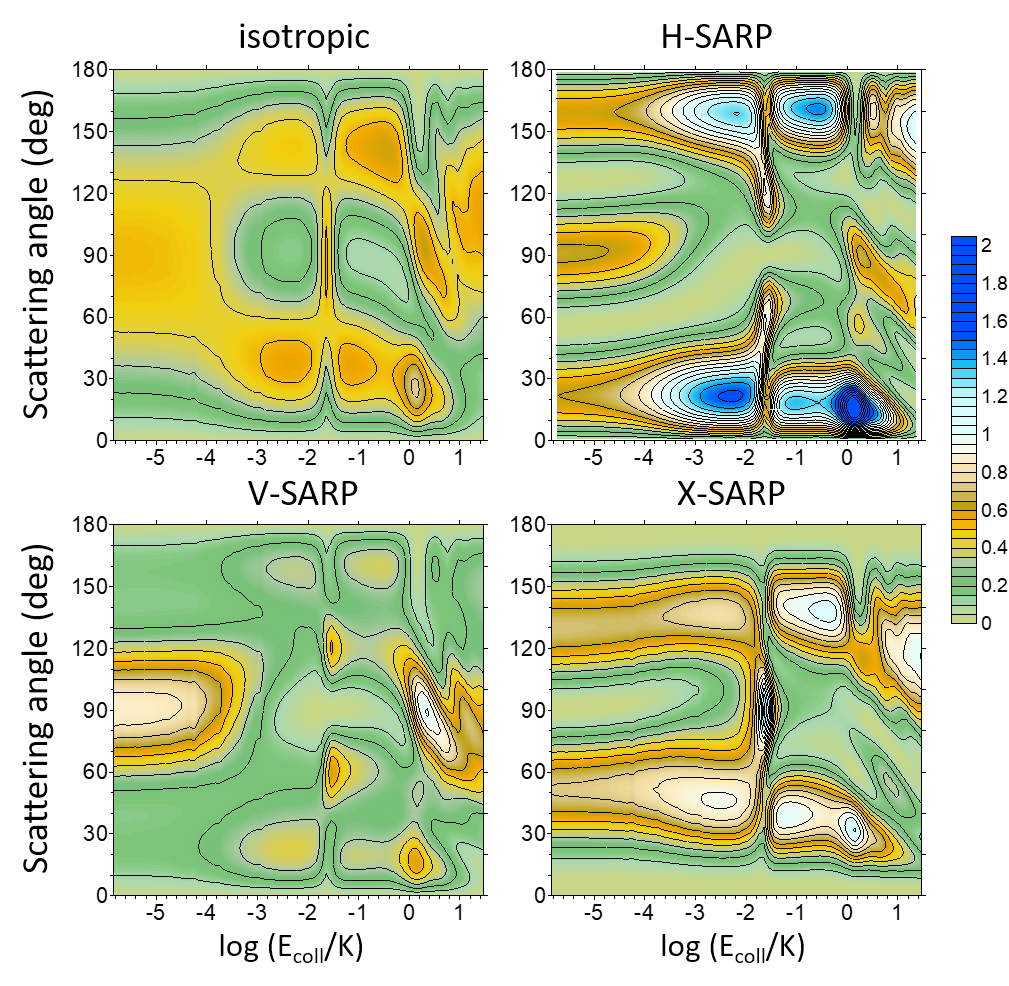}
\end{center}
\caption{Normalized Differential Cross Sections ($I(\theta)$ divided by the ICS) for the D$_2$ ($v=2,j=2 \to v'=2,j'=0$) by collisions with He as a function of the collision energy for different initial preparations of the D$_2$ rotational state. For a given collision energy, the normalized DCS is calculated as the DCS divided by the ICS at that given energy, and it is multiplied by  $\sin\theta$.
}
\label{fig:DCS_3d}
\end{figure}

\begin{figure}[h!]
\begin{center}
\includegraphics[width=1.0\linewidth]{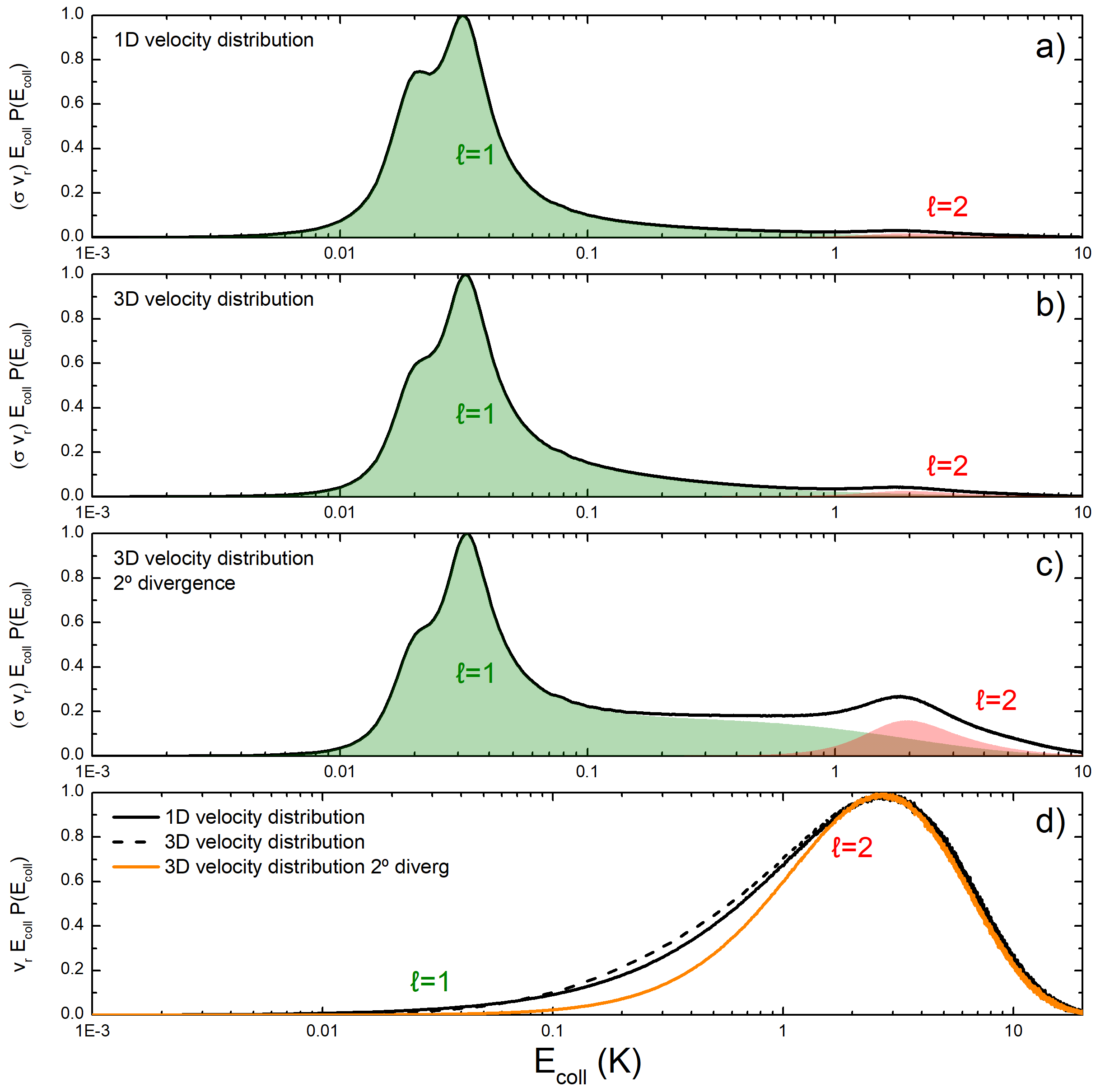}
\end{center}
\caption{Energy dependent rate coefficients multiplied by the experimental velocity distribution assuming a 1D distribution (panel a), a 3D distribution (panel b), and a 3D distribution with a large beam divergence (panel c). The rate coefficients are multiplied by E$_{\rm coll}$, so its area in a logarithm $E_{\rm coll}$ scale is proportional to the flux. Values are renormalized so its maximum value is 1. Contributions of $\ell$=1, and 2 are highlighted in olive, and red. The three aforementioned velocities distributions are shown in panel d, also multiplied by $E_{\rm coll}$, so the area is proportional to the incoming flux at a given energy in a logarithm scale.
}
\label{fig:flux}
\end{figure}

\begin{figure}[h!]
  \centering
  \includegraphics[width=1.0\linewidth]{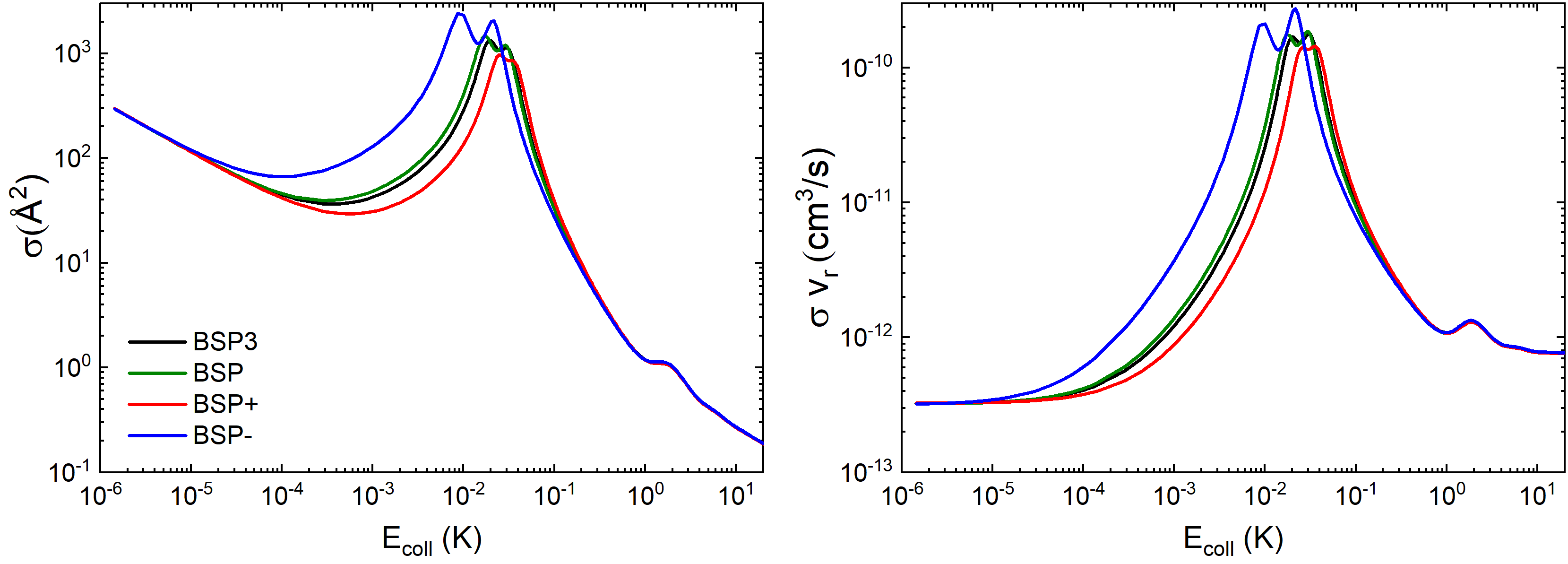}
  \caption{Left panel: Excitation functions for He+D$_2 (v=2,j=2 \to
  v'=2,j'=0$) collisions calculated for different PESs: BSP3~\cite{BSP3} (that used throughout the manuscript), BSP (an earlier version of the PES) and BSP- and BSP+,  which are fitted to (energy+uncertainty) and
 (energy-uncertainty) of the ab initio data, respectively, and hence are the ``upper limit'' and ``lower limit''
 versions of the BSP potentials. Right panel:Energy dependent rate coefficients for the different PESs.}
  \label{fig:BSP}
\end{figure}

\begin{figure}[h!]
\begin{center}
\includegraphics[width=1.0\linewidth]{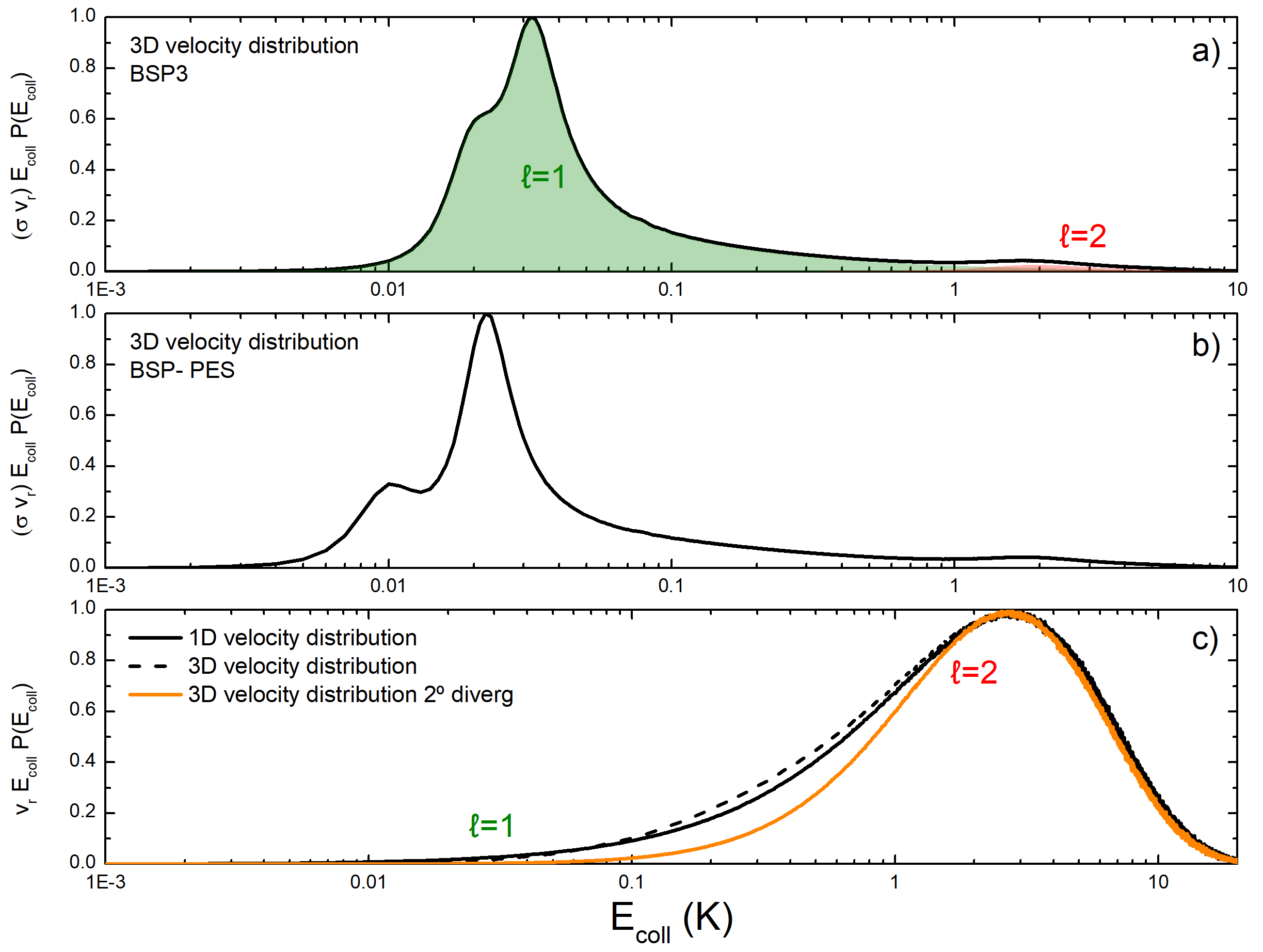}
\end{center}
\caption{As Fig \ref{fig:flux} but showing the comparison between the flux obtained using BSP3 or BSP- PES. Although for BSP- the resonance is shifted towards lower energies, $\ell$=1 is still dominant when integration over the velocity distribution is carried out.
}
\label{fig:fluxmod}
\end{figure}

\begin{figure}[h!]
  \centering
  \includegraphics[width=1.0\linewidth]{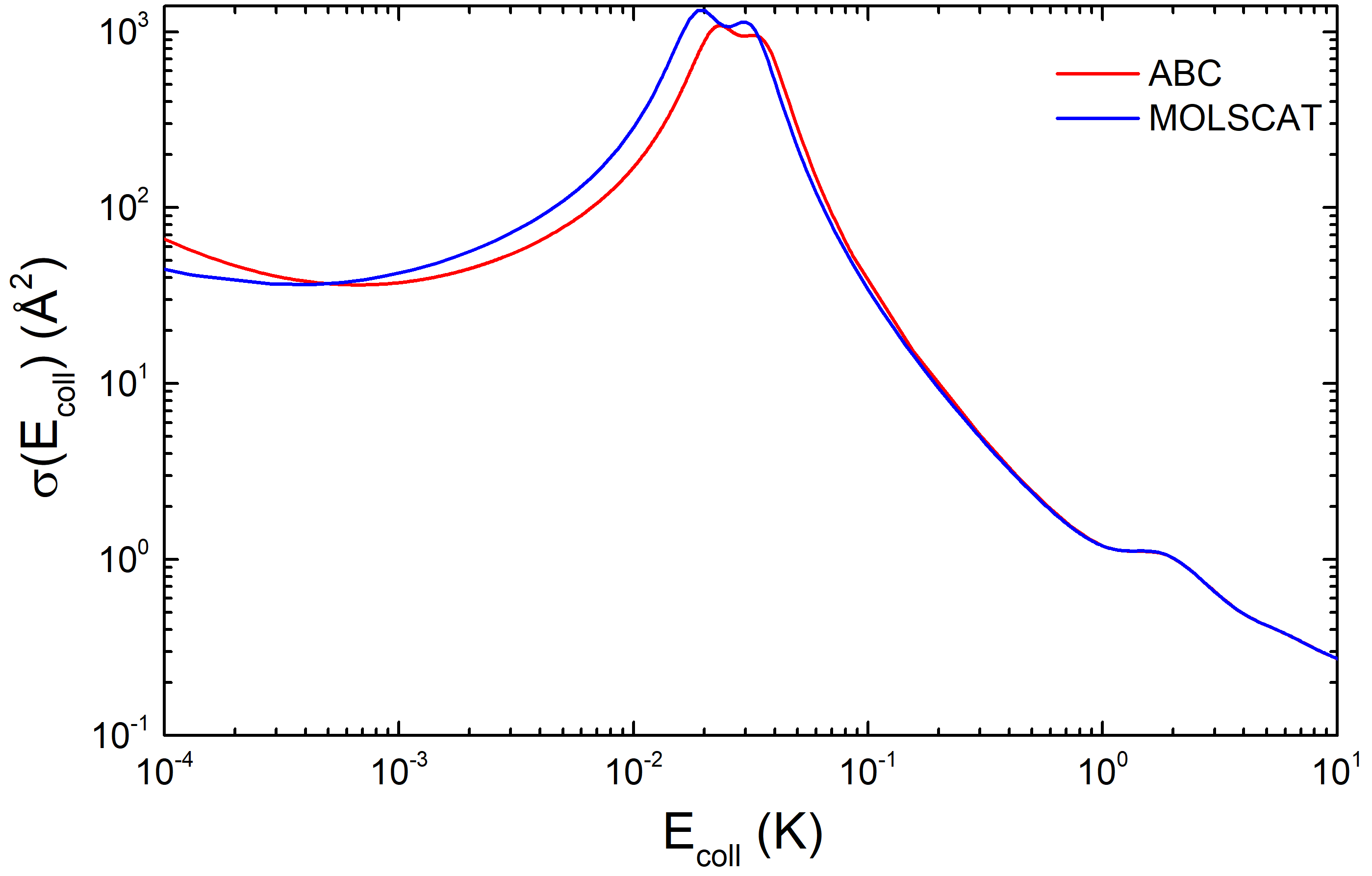}
  \caption{Comparison between the Excitation function calculated using either MOLSCAT (solid blue line) or ABC (solid red line), which uses hyperspherical coordinates for the scattering.}
  \label{fig:molscatvsabc}
\end{figure}

\begin{figure}[h!]
  \centering
  \includegraphics[width=1.0\linewidth]{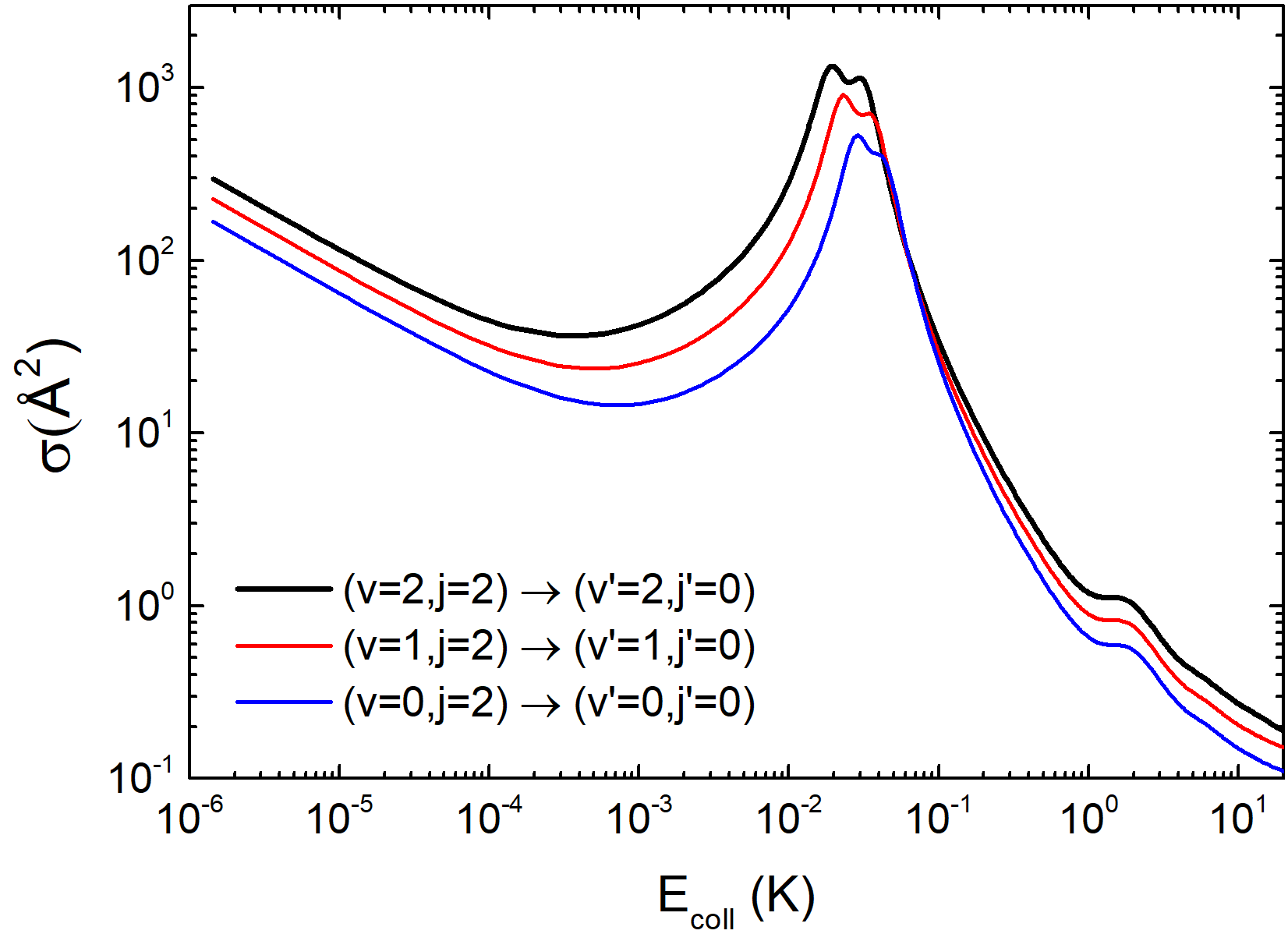}
  \caption{Excitation functions for He+D$_2 (v,j=2 \to
  v',j'=0$) collisions calculated for different D$_2$ vibrational levels.}
  \label{fig:icsvj2}
\end{figure}

\clearpage

\bibliography{cite}